\documentclass[
aps,
superscriptaddress,nofootinbib,
reprint,
]{revtex4-2}

\usepackage{xcolor}
\usepackage{graphicx}
\usepackage{amsmath,amssymb,amsfonts,mathrsfs,bm}
\usepackage[
colorlinks=true,
linkcolor=blue,
citecolor=blue,
]{hyperref}

\def\be#1\ee{\begin{equation}#1\end{equation}}
\def\bl#1\el{\begin{align}#1\end{align}}
\def\ba#1\ea{\begin{align*}#1\end{align*}}
\def\ddd{\mathrm{d}}
\def\l{\left}
\def\r{\right}
\def\nn{\nonumber}

\begin{document}

\title{Constraining matter-bounce scenario from scalar-induced vector perturbations}

\begin{abstract}
    Bouncing cosmologies, while offering a compelling alternative to inflationary models, face challenges from the growth of vector perturbations during the contracting phase. While linear vector instabilities can be avoided with specific initial conditions or the absence of vector degrees of freedom, we demonstrate the significant role of secondary vector perturbations generated by nonlinear interactions with scalar fluctuations. Our analysis reveals that in a broad class of single-field matter-bounce scenarios, these secondary vector perturbations get unacceptably large amplitudes, provided the curvature fluctuations are consistent with cosmic microwave background observations. This finding underscores the crucial importance of scalar-induced vector perturbations in bouncing cosmology and highlights the need for further investigation into their potential impact on the viability of these models.
\end{abstract}

\author{Mian Zhu}
\email[]{mian.zhu@uj.edu.pl}
\affiliation{ College of Physics, Sichuan University, Chengdu 610065, China}
\affiliation{Faculty of Physics, Astronomy and Applied Computer Science,
	Jagiellonian University, 30-348 Krakow, Poland}

\author{Chao Chen}
\email[Corresponding author:~]{chaochen012@gmail.com}
\affiliation{Department of Physics, School of Science, Jiangsu University of Science and Technology, Zhenjiang, 212003, China}
\affiliation{Jockey Club Institute for Advanced Study, The Hong Kong University of Science and Technology, Hong Kong, China}

\maketitle

\section{Introduction}

Inflation~\cite{Guth:1980zm}, the standard paradigm of the early-universe cosmology, provides a natural way to explain the formation of large-scale structures (LSS) and the observation of the cosmic microwave background (CMB). Nonetheless, inflationary cosmology may suffer from the initial singularity problem \cite{Borde:1993xh,Borde:2001nh, Lesnefsky:2022fen,Geshnizjani:2023edw} and the trans-Planckian problem \cite{Martin:2000xs,Bedroya:2019snp,Cai:2019hge}. These challenges motivate us to explore alternative early universe scenarios such as the nonsingular bouncing cosmology \cite{Novello:2008ra,Lehners:2008vx,Brandenberger:2016vhg,Cai:2016hea}, where a contraction phase takes place prior to the expansion phase.  
While bouncing cosmology offers an intriguing alternative for the early universe, it faces significant challenges. Conceptual issues~\cite{Brandenberger:2009jq,Ijjas:2018qbo} and its compatibility with CMB observations~\cite{Cai:2014xxa,Cai:2014bea} remain critical concerns. There are also extensive debates surrounding specific problems of bouncing cosmologies ~\cite{Karouby:2010wt,Karouby:2011wj, Bhattacharya:2013ut,Cai:2013vm, Grain:2020wro,Cline:2003gs,Vikman:2004dc,Xia:2007km,Easson:2016klq, Libanov:2016kfc,Kobayashi:2016xpl,Akama:2017jsa,Cai:2009fn, Gao:2014hea,Gao:2014eaa, Quintin:2015rta,Li:2016xjb,Ageeva:2021yik,Ageeva:2022fyq,Akama:2022usl} and proposed solutions~\cite{Khoury:2001wf,Middleton:2008rh,Lin:2017fec,Cai:2016thi, Creminelli:2016zwa,Cai:2017tku,Cai:2017dyi,Kolevatov:2017voe,Akama:2018cqv,Mironov:2019qjt,Akama:2019qeh,Ilyas:2020qja,Zhu:2021whu,Cai:2022ori}; a comprehensive review of these challenges is available in Ref.~\cite{Battefeld:2014uga}.

In this paper, we highlight another challenge for bouncing cosmology, the overproduction of vector perturbations, a problem overlooked in the community. Early studies~\cite{Battefeld:2004cd,Bojowald:2007hv} demonstrated that linear vector perturbations scale as $S_i(k) \propto a^{-2}$, leading to its growth that can break down the perturbation theory. Resolving this issue typically requires specific model constructions or finely tuned initial conditions for vector perturbations. For instance, a single-field bouncing scenario lacks vector degrees of freedom, preventing primordial vector fluctuations from vacuum fluctuations.

However, secondary vector perturbations inevitably arise from nonlinear interactions with primordial curvature fluctuations $\zeta$. Those fluctuations cannot be arbitrarily fine-tuned, as the power spectrum of curvature fluctuation $\mathcal{P}_{\zeta}$ is determined by CMB observations. In Ref.~\cite{Mena:2007ve}, scalar-induced vector perturbations (SIVPs) are investigated in specific collapsing universes with theoretical considerations. For the first time, we in this paper connect the power spectrum $\mathcal{P}_{\zeta}$ to CMB observations, establishing a lower bound for the energy density of SIVPs. Specifically, we work in a matter-bounce scenario~\cite{Brandenberger:2012zb}, a simple-yet-significant bouncing scenario where  nearly scale-invariant curvature fluctuation is generated in a matter-dominated contraction phase (i.e., the effective equation-of-state parameter is zero). Our results demonstrate that the energy density of SIVPs becomes comparable to the background energy density at the end of the matter-contraction phase, provided the contraction is driven by a k-essence scalar field. This significant backreaction poses a serious challenge to the viability of the matter-bounce scenario.

The paper is organized as follows. We introduce matter-bounce cosmology in Sec. II, focusing on the scalar fluctuations. We discuss the gauge dependence problem of secondary vector perturbations in Sec. III, and provide our resolution. Section IV is devoted to the formalism of scalar-induced vector fluctuations in bouncing cosmology. After getting the analytical expressions of SIVPs in Sec. IV, we compute the energy density of SIVPs in Sec. V and show that it greatly exceeds the background energy density, indicating the potential breakdown of cosmological perturbation theory. We draw our conclusion in Sec. VI and discuss future extensions of our work.

\section{Scalar Perturbations in Bounce Cosmology}

We first review the curvature fluctuation in matter-bounce cosmology. We will adopt the uniform field gauge where $\phi = \phi(t)$. The uniform field gauge finds its broad application in the computation of cosmological perturbations, enabling us to use the result of~\cite{Gao:2011qe}. Specifically, the perturbed metric is written as
\begin{widetext}
	\begin{align}
		ds^2 = a^2 \Big[ -e^{2\alpha} d\tau^2 + e^{2\zeta} \delta_{ij} (dx^i + e^{-2\zeta} \partial^i \beta d\tau) (dx^j + e^{-2\zeta} \partial^j \beta d\tau ) \Big] ~,
	\end{align}
\end{widetext}
where $d\tau = dt/a$ is the conformal time, and $\zeta$ is the comoving curvature perturbation. The parameters $\alpha$ and $\beta$ can be solved from constraints. We set the Planck mass $M_p = 1$, and a prime to denote differentiation with respect to $\tau$, unless specified. In matter-bounce, the scalar factor scales as $a \propto \tau^2$ and can be parametrized as
\begin{equation}
	\label{eq:bouncebg}
	a(\tau) = (\tau /\tau_0)^2 ~,~
	\tau < \tau_0 < 0 ~,
\end{equation}
where $\tau_0$ labels the end of the contraction phase. It will also be useful to define a comoving Hubble parameter $\mathcal{H} \equiv a^{\prime}/a = 2/\tau < 0$. The background energy density is given by the Friedmann's equation  $\rho_{\rm bg}(\tau) = 3H^2 = 12 \tau_0^4/\tau^6$.
In the framework of k-essence theory \cite{Armendariz-Picon:2000ulo}, 
\begin{equation}
	\label{eq:kessence}
	S = \int d^4x \sqrt{-g} \left[ \frac{R}{2} + K\left( \phi, X \right) \right] ~,~
	X \equiv -\frac{\partial_{\mu} \phi \partial^{\mu} \phi}{2}  ~,
\end{equation}
it has the following form:
\begin{align}
	\label{eq:SS2}
	S_S^{(2)} &= \int d\tau d^3x a^2 \Big[ -3 \zeta^{\prime 2} + (\partial_i \zeta)^2 -3 \mathcal{H}^2 \alpha^2 \left( 1 - \frac{\epsilon}{3c_s^2} \right) 
	\nn\\& + 2\partial \alpha \partial \zeta + 6\mathcal{H} \alpha \zeta^{\prime} + 2(\zeta^{\prime} - \alpha \mathcal{H}) \partial_i^2 \beta \Big] ~.
\end{align}
Variation with respect to $\alpha$ and $\beta$ gives the constraint equations
\begin{equation}
	\alpha = \frac{\zeta^{\prime}}{\mathcal{H}} ~,~ \beta = - \frac{\zeta}{\mathcal{H}} + \frac{3}{2c_s^2} \partial^{-2} (\zeta^{\prime}) ~,
\end{equation}
where we used $\epsilon \equiv -\dot{H}/H^2 = 3/2$ in the matter-contraction phase. The quadratic action for curvature perturbation with the help of constraint equations becomes~\cite{Garriga:1999vw,Chen:2010xka,Gao:2011qe}
\begin{equation}
    S_{\zeta}^{(2)} = \int d\tau d^3x \frac{z_s^2}{2} \l[ \zeta^{\prime 2} - c_s^2 (\partial_i \zeta)^2 \r] ~,~  z_s^2 = \frac{3a^2}{c_s^2} ~,
\end{equation}
where we used the fact that the effective slow-roll parameter $\epsilon \equiv -\dot{H}/H^2 = 3/2$ in the matter-contraction phase, and we regarded the sound speed for curvature perturbations, $c_s^2 \equiv K_{,X}/(K_{,X} + 2XK_{,XX})$, as a constant for simplicity.
Working in the Fourier space with a canonically normalized mode function $v_k = z_s \zeta_k$, $z_s \equiv \sqrt{3}a/c_s$, the dynamical equation for curvature perturbations becomes
\begin{equation}
	v_k^{\prime \prime} + \left( c_s^2k^2 - \frac{2}{\tau^2} \right) v_k = 0 ~.
\end{equation}
Imposing the vacuum initial condition, we get the expression for curvature fluctuations as~\cite{Allen:2004vz,Vitenti:2011yc,Durrer:1993tti,Pinto-Neto:2013zya,Ota:2022xni,Bardeen:1980kt, Brandenberger:1983tg,Cai:2008qw,Hwang:2017oxa,Domenech:2020xin,Domenech:2021ztg,Giovannini:1997gp,Acquaviva:2002ud,Bartolo:2004if,Chang:2022dhh,Zeldovich:1971mw,Parker:1974qw,Ye:2022tgs,DeLuca:2019ufz,Inomata:2019yww}
\begin{equation}
	\label{eq:zetak}
	\zeta_k (\tau) \equiv \frac{v_k(\tau)}{z_s} = \frac{e^{-ik c_s \tau} c_s}{\sqrt{6c_sk}} \left( 1 - \frac{i}{c_s k \tau} \right) \left( \frac{\tau_0}{\tau} \right)^2 ~.
\end{equation}
In contrast to the vanilla slow-roll inflation case, the curvature perturbations grow on superhorizon scales $|k\tau| \ll 1$ in the matter-contraction phase (see e.g., Ref.~\cite{Weinberg:2008zzc}). Hence, one needs to evaluate the curvature power spectrum at the end of the contraction phase:
\begin{align}
    \langle \zeta_{\vec{k}} \zeta_{\vec{p}} \rangle (\tau = \tau_0) 
    &= (2\pi)^3 \delta(\vec{k} + \vec{p}) |\zeta_{\vec{k}}|^2 
    \nn\\&= (2\pi)^3 \delta(\vec{k} + \vec{p}) \frac{c_s}{6k} \left( 1 + \frac{1}{k^2 c_s^2 \tau_0^2} \right) ~.    
\end{align}
From the definition of the scalar power spectrum,
\begin{equation}
\label{eq:Pzetadef}
     \langle \zeta_{\vec{k}} \zeta_{\vec{p}} \rangle = (2\pi)^3 \delta(\vec{k} + \vec{p}) \frac{2\pi^2}{k^3} \mathcal{P}_{\zeta}(k) ~,
\end{equation}
we derive
\begin{equation}
\label{eq:Pzeta}
    \mathcal{P}_{\zeta}(k,\tau_0) = \frac{k^2 c_s}{12\pi^2} \left( 1 + \frac{1}{k^2 c_s^2 \tau_0^2} \right) \simeq \frac{1}{12\pi^2 c_s \tau_0^2} ~,
\end{equation}
which is scale invariant on superhorizon scales.

\section{Vector Perturbations in Bounce Cosmology}

In an FLRW universe, the most general perturbed metric, including only vector perturbation, is given by~\cite{Mukhanov:1990me}
\begin{equation}
	d s^2 = a^2(\tau) \l[ -d \tau^2 - 2G_i d\tau d x^i + (\delta_{ij} + F_{ij}) d x^i d x^j \r] ~,
\end{equation}
where $F_{ij} \equiv \partial_i F_j + \partial_j F_i$, $\partial_i F^i = 0$. A gauge-invariant quantity can be defined as $\sigma_i = G_i + {F}_i^{\prime}$ (note that our definition of $G_i$ and $F_i$ is for the conformal observer, slightly different from the classical literatures, e.g., \cite{Durrer:1993tti}). 
In the following, we will verify that different gauge choices of secondary vector perturbation lead to the same Einstein equations in our scenario, and we can safely adopt either $F_i = 0$ or $G_i = 0$ for convenience.

\subsection{Problem of gauge choice}

There are two problems of gauge choice in the computation of SIVPs in the matter-bounce scenario. First, Ref.~\cite{Allen:2004vz} points out that, throughout the bounce, the rapid growth of the Bardeen potential can lead to the invalidation of the Newtonian gauge. This problem is revisited in \cite{Vitenti:2011yc, Pinto-Neto:2013zya}, and the linearity conditions are proposed to examine whether a specific gauge choice is problematic in the bouncing scenario.

To illustrate the problem, we write down the metric in the ADM form
\begin{equation}
	ds^2 = -N^2 d\tau^2 + \gamma_{ij} (N^i d\tau + dx^i) (N^j d\tau + dx^j) ~,
\end{equation}
and a perturbed metric on a flat FLRW background on a linear level can be represented by
\begin{equation}
	\label{eq:scalarfluctuationfull}
	N = a e^{\varphi} ~,~ N_i = a^2 \partial_i \mathcal{B} ~,~ \gamma_{ij} = a^2 e^{2\psi} e^{- 2\partial_i\partial_j \mathcal{E}} ~,
\end{equation}
where $\partial_i$ represents the covariant derivative with respect to the three-subspace. Here for simplicity we only included the scalar fluctuations. For the particular exponential form of perturbations, see Sec. IV of Ref. \cite{Ota:2022xni} for more details. The exponential form is to be understood by the following formula:
\begin{equation}
	e^{h_{ij}} = \delta_{ij} + h_{ij} + \frac{1}{2} h_{ik}h^{kj} + \mathcal{O}(h^3) ~.
\end{equation}

Specific gauge choices can be made to ensure that there is only one scalar degree of freedom in our setup. The Newtonian gauge is 
\begin{equation}
	\mathcal{B} = \mathcal{E} = 0 ~;~ \varphi = \Phi = - \psi ~.
\end{equation}

The linear perturbation theory remains valid only if 
\begin{equation}
	\mathcal{B} \ll 1 ~,~ \mathcal{E} \ll 1 ~,~ \varphi \ll 1 ~,~ \psi \ll 1 ~.
\end{equation}
In addition, the perturbed quantities, namely the Einstein equations, expansion rate, curvature scalar, etc., must remain small compared to the background ones. Those conditions are concluded in Ref. \cite{Vitenti:2011yc} and presented in a gauge-invariant manner as in Eqs. (62)--(65) therein:
\begin{equation}
	\label{eq:geoperteq1}
	\left| \frac{a \delta \Theta}{\mathcal{H}} \right| \ll 1 ~;~ a \delta \Theta \equiv - \partial^2 (\mathcal{E}^{\prime} - \mathcal{B}) + 3 (\mathcal{H}\varphi + \psi^{\prime}) ~,
\end{equation}
\begin{equation}
	\label{eq:geoperteq2}
	\left| \frac{\partial^2 \psi}{\mathcal{H}^2} \right| \ll 1 ~,~ \left| \frac{\partial^2 \psi}{\mathcal{H}^2 + 2\mathcal{H}^{\prime}} \right| \ll 1 ~,
\end{equation}
\begin{equation}
	\label{eq:geoperteq3}
	\left| \frac{\partial^2 \delta \sigma}{a\mathcal{H}} \right| \ll 1 ~,~ \left| \frac{\partial^2 \varphi}{\mathcal{H}^{\prime} - \mathcal{H}^2} \right| \ll 1 ~.~ \delta \sigma \equiv -a (\mathcal{E}^{\prime} - \mathcal{B}) ~.
\end{equation}
Notice that the condition on matter-sector $|\delta \rho / \rho| \ll 1$ and $|\delta p / p| \ll 1$ would imply weaker constraints compared to the above geometric ones, so we do not present their explicit form here. 

Now we see that in Newtonian gauge the condition \eqref{eq:geoperteq1} simplifies to $|\Phi - \Phi^{\prime}/\mathcal{H} | \ll 1$. The Bardeen potential $\Phi$ is related to the curvature fluctuation via \cite{Bardeen:1980kt,Brandenberger:1983tg}
\begin{equation}
	\zeta = \Phi + \frac{\mathcal{H}}{\mathcal{H}^2 - \mathcal{H}^{\prime}} ( \Phi^{\prime} + \mathcal{H} \Phi ) ~.
\end{equation}
In matter-contraction phase it reduces to 
\begin{equation}
	\zeta = \frac{5}{3} \Phi + \frac{\tau}{3} \Phi^{\prime} ~.
\end{equation}
Using \eqref{eq:zetak} we get $\Phi$ in the Fourier domain to be
\begin{equation}
	\Phi_k = \sqrt{\frac{3c_s}{2k}} \left( \frac{\tau_0}{\tau} \right)^2 e^{-ic_sk\tau} \frac{-3i + 3c_sk\tau + i c_s^2 k^2 \tau^2}{(c_sk\tau)^3} ~.
\end{equation}
Thus, in the superhorizon limit one has $|\zeta_k / \Phi_k| \propto |k\tau|^2 \ll 1$. The smallness of $\zeta_k$ does not necessarily lead to $|\Phi| \ll 1$. Once $\Phi$ reaches order unity, the linearity conditions are not fulfilled and the perturbation theory can be questionable. 

Thus, to calculate SIVPs, we should work with specific gauges that satisfy the linearity conditions. In matter-bounce scenario, there is an additional problem. If the matter-contraction phase is governed by a pressureless dust, then the latter condition of \eqref{eq:geoperteq2} can never be satisfied since $\mathcal{H}^2 + 2\mathcal{H}^{\prime} = 0$ as a result of the vanishing background pressure. Additionally, we will have $c_s^2 = w = 0$, which can potentially lead to the strong coupling problem. Thus, we set up our formalism in the context of k-essence theory, and the matter-contraction phase shall be understood in a time-averaged sense, $\langle w \rangle = 0$, resulting from the Virial theorem; see, e.g., Ref. \cite{Cai:2008qw}. In this sense, we can safely adopt a nonzero $c_s^2$ without worrying about the breakdown of the $c_s^2 = w$ condition. Similarly, the latter constraint of \eqref{eq:geoperteq2} now represents $\delta \mathcal{R} / \langle G_i^i \rangle = 0$ with $\mathcal{R}$ the curvature scalar, which is not applicable again. The correct condition is $\delta \mathcal{R} / G_i^i = 0$, and since $G_i^i = p$ and $G_i^i = \rho$ are of the similar order except for discrete time slices, this condition simply degenerates with the $\delta \mathcal{R} / G_0^0 = 0$ condition, namely the first constraint in \eqref{eq:geoperteq2}.

It is straightforward to find 
\begin{equation}
	\zeta_k^{\prime} \simeq -3\zeta_k/\tau \simeq -3\zeta_k \mathcal{H}/2 ~,
\end{equation}
for superhorizon modes. Thus, $\alpha$ would be proportional to $\zeta$ for those modes, and its smallness is guaranteed since $\zeta^2 \sim \mathcal{O}(10^{-9})$ from the CMB observations. Intuitively, the violation of the linearity condition in Newtonian gauge comes from the fact that the Bardeen potential $\Phi$ grows faster than the curvature fluctuation. Gauges that ensure the perturbed geometric quantities to grow not faster than $\zeta$ could be natural solutions to the problem. The smallness of $\beta$ can also follow directly due to the smallness of $\alpha$ and $\zeta$. One may also use the facts $\varphi = \alpha$, $\psi = \zeta$ and $\mathcal{B} = e^{-2\zeta} \beta$ to verify that all conditions \eqref{eq:geoperteq1}, \eqref{eq:geoperteq2} and \eqref{eq:geoperteq3} are fulfilled. Thus, we confirm that the uniform field gauge is a good choice for both the perturbative computation and the requirement from linearity condition (except for the $\delta \mathcal{R}/G_i^i$ one, which we explain above). 

Now we come to the second puzzle. It is well known that the gravitational wave power spectrum of scalar-induced gravitational waves (SIGWs) generically depends on the gauge choice of linear scalar perturbation \cite{Hwang:2017oxa}. Specifically, in Ref. \cite{Domenech:2020xin} it is shown that deep inside the horizon several gauges including the Newtonian gauge are robust for the study of SIGWs and give almost identical results, while the uniform field gauge may not be a good candidate. Additionally, the gauge problem in the SIGW is not totally resolved, since it is not clear which gauge is most suitable for the SIGW when connecting to observational GWs.

\subsection{Our resolution of the gauge dependence problem}

In the study of SIGWs, two main approaches exist to settle the gauge dependence issue. The first relies on constructing a gauge-invariant formulation of tensor modes, while the latter focuses on finding an appropriate gauge choice that best describes the GW detection. Each approach has its merits and drawbacks. The former approach leads to gauge-invariant results, but there is no clear connection between the gauge-invariant variable and the real observable. For example, both the Bardeen potential $\Phi$ and the curvature fluctuations are all gauge-invariant variables of scalar modes, but they are quite different, as analyzed in the above section. Ultimately, choosing a particular gauge-invariant combination is no different than selecting a specific gauge. See discussions in Ref. \cite{Domenech:2020xin}. The latter approach, associated with an apparent gauge dependent, can have a clear physical motivation. For instance, it is argued that the most suitable gauge for the study of GWs is the gauge where the coordinates follow a geodesic congruence, e.g. a frame where the mirrors of the interferometer are fixed, also known as synchronous gauge \cite{DeLuca:2019ufz}.

The gauge dependence of secondary fluctuations is rooted in the fact that scalar, vector, and tensor modes mix with each other. It is one of the most intriguing questions in the study of secondary fluctuations that remains to be resolved. We will reduce the gauge issue of the SIVPs to a minimum and show that the instability of SIVPs indeed takes place by proceeding with different gauge choices, similar to the approaches in the study of SIGWs. Specifically, we work out that the instability of SIVPs appears in two gauges. The uniform field gauge (UFG) is our first choice since the perturbative calculations in Horndeski theory are mainly performed in this gauge \cite{DeFelice:2011uc, Kobayashi:2011nu}. If the instability of SIVPs appears in the UFG in matter-bounce scenario, then any result of matter-bounce that relies on the perturbative calculations in UFG becomes unconvincing. Another gauge we will adopt is the constant curvature gauge (CCG), which is robust in the matter-bounce scenario, while other gauges such as the Newtonian gauge fail to satisfy the linearity condition~\cite{Vitenti:2011yc}. We find that SIVPs are overproduced in both gauges.

\section{Scalar-Induced Vector Perturbations in Bounce Cosmology}

In the FLRW universe, the most general perturbed metric, including only vector perturbation, is \cite{Mukhanov:1990me}
\begin{equation}
    ds^2 = a^2(\tau) \l[ -d \tau^2 - 2G_i d\tau d x^i + e^{F_{ij}} d x^i d x^j \r] ~,
\end{equation}
where $F_{ij}$ satisfies $F_{ij} = \partial_i F_j + \partial_j F_i$ and $\partial^i F_i = 0$, and $G_i$ is divergent free, $\partial_i G^i = 0$. Since there is no vector degree of freedom in our setup, $F_i$ and $G_i$ should be regarded as second-order fluctuations induced by first-order perturbations. The Einstein equation at second order involving the vector fluctuations has the following form 
\begin{equation} 
\label{eq:vectordynamicaleq}
G^{\lambda \prime}(\tau,\vec{k}) + 2 \mathcal{H} G^{\lambda}(\tau,\vec{k}) = S^{\lambda}(\tau,\vec{k}) ~,
\end{equation}
where $G^{\lambda}$ is the secondary vector fluctuation in Fourier space and the form of $S^{\lambda}$ remains to be decided after the gauge choice of curvature fluctuation. Since the spectrum of scalar-induced gravitational waves is dependent on the gauge choice of scalar fluctuation, we will work with both the uniform field gauge and the constant curvature gauge.

The vector power spectrum is defined as
\begin{equation}
	\langle G^{\lambda} (\vec{k}) G^{s} (\vec{p}) \rangle \equiv  (2\pi)^3 \delta(\vec{k}+ \vec{p}) \delta^{\lambda s}  \frac{2\pi^2}{k^3} \mathcal{P}_G (\tau,\vec{k}) ~.
\end{equation} 
Using \eqref{eq:vectordynamicaleq}, the vector power spectrum is computed as
\begin{align} 
	\label{eq:PG}
	& \quad \mathcal{P}_G(\tau, k)  \nonumber = \int_{\frac{1}{\sqrt{2}}}^{\infty} dt \int_{-\frac{1}{\sqrt{2}}}^{\frac{1}{\sqrt{2}}} ds \frac{(1-2s^2)(2t^2-1)(2st+1)^2}{4(t^2-s^2)^2 } \\ 
	& \times
	\mathcal{P}_{\zeta}\left(\frac{k(t-s)}{\sqrt{2}},\tau \right) \mathcal{P}_{\zeta} \left(\frac{k(t+s)}{\sqrt{2}}, \tau \right)  \left|  \mathcal{I}(t,s,z) \right|^2 ~,
\end{align}
where $z \equiv k\tau$.

\subsection{SIVPs in the uniform field gauge}

The second-order vector $G_i^{(2)}$ is determined by the $ij$ components of the Einstein equations,
\begin{equation}
	G_{ij}^{(2)} = T_{ij}^{(2)} ~,~ i \neq j ~.
\end{equation}
Although it is possible to derive the secondary vector fluctuations using momentum constraints (as there are no vector degrees of freedom in our specific scenario), the method of computing with Einstein equations will prove useful in future studies involving vector fields. Examples include primordial magnetogenesis~\cite{Kandus:2010nw} and baryon asymmetry~\cite{Giovannini:1997gp}. For detailed explanations of both methods and their equivalence, see Refs.~\cite{Acquaviva:2002ud,Bartolo:2004if}.

The perturbed metric with unitary gauge can be written as
\begin{widetext}
	\begin{equation}
		\label{eq:metricunitaryvec}
		ds^2 = a^2(\tau) \left[ -e^{2\alpha} d\tau^2 + e^{2\zeta} e^{F_{ij}} (dx^i + e^{-2\zeta} e^{F^{ik}} (\partial_k \beta - G_k) d\tau) (dx^j + e^{-2\zeta}e^{-F^{jl}} (\partial_l \beta - G_l) d\tau ) \right] ~,
	\end{equation}
\end{widetext}
with $F_{ij} \equiv \partial_i F_j + \partial_j F_i$, $\partial_i F^i = 0$. In terms of the ADM variables,
\begin{equation}
	N = ae^{\alpha} ~,~ N_i = a^2 (\partial_i \beta - G_i) ~,~ \gamma_{ij} = a^2 e^{2\zeta} e^{F_{ij}} ~,
\end{equation}
the inverse metric is
\begin{align}
		&g^{00} = - \frac{1}{N^2} = -a^{-2}e^{-2\alpha} ~,
		\\&
		g^{0i} = \frac{N^i}{N^2} = a^{-2} e^{-2(\alpha + \zeta)} e^{-F^{ij}}  (\partial_j \beta - G_j)  ~,
\end{align}
\begin{align}
	g^{ij} &= \frac{e^{-2\zeta}}{a^2}  \Big[ e^{-F^{ij}} 
	\nn\\&- e^{-2(\alpha + \zeta) } e^{-F^{ik}} e^{-F^{jl}} (\partial_k \beta - G_k) (\partial_l \beta - G_l) \Big] ~.
\end{align}
We remind the readers that $\alpha$ and $\beta$ are merely Lagrangian multipliers and can be solved through the constraint equations:
\begin{equation}
	\alpha = \frac{\zeta^{\prime}}{\mathcal{H}} ~,~ \beta = - \frac{\zeta}{\mathcal{H}} + \frac{3}{2c_s^2} \partial^{-2} (\zeta^{\prime}) ~.
\end{equation}

The computation of geometric quantities based on the metric perturbations in Eq. \eqref{eq:metricunitaryvec} is straightforward. We will first work in the vector gauge $F_i = 0$. Later on, we work in another gauge $G_i = 0$ and confirm that both choices give identical results. We present some useful expressions below (utilizing the identity $\mathcal{H}^{\prime} = -\mathcal{H}^2/2$, applicable in the matter-bounce scenario, and the notation $\partial^2 \equiv \partial_i \partial^i$),
\begin{equation}
	\label{eq:intrinsic}
	G_0^{(0) 0} = -3 \frac{\mathcal{H}^2}{a^2} ~,
\end{equation}

\begin{align}
	\label{eq:Gij2}
	&G_{ij}^{(2)} + \frac{1}{2} \delta_{ij} \mathcal{P} 
	\nn\\& = \frac{1}{2} \partial_i G_j^{\prime} + \mathcal{H} \partial_i G_j - \frac{1}{2} \partial_i \alpha \partial_j \alpha + \frac{1}{2} \partial_i \zeta \partial_j \zeta 
	\nn\\&+ ( \partial_i \beta^{\prime} + \partial_i \alpha + 2 \mathcal{H} \partial_i \beta  ) \partial_j \zeta + \partial_i \beta \partial_j \zeta^{\prime} 
	\nn\\& + \alpha \partial_i \partial_j ( \beta^{\prime} +  2 \mathcal{H} \partial_i \beta) + \frac{1}{2} (\partial^2 \beta + \alpha^{\prime} - \zeta^{\prime}) \partial_i \partial_j \beta 
	\nn\\&- \frac{1}{2} \partial^a \beta \partial_a \partial_i \partial_j \beta  + ( i \longleftrightarrow j)  ~,
\end{align}
where $\mathcal{P}$ represents the symmetric part of $G_{ij}^{(2)}$. Later on, we will see that the SIVPs are independent of $\mathcal{P}$, freeing us from computing with its tediously long expressions.

For the matter sector, we obtain
\begin{equation}
	\label{eq:Tmunugeneral}
	T_{\mu \nu} = ( \rho + P ) u_\mu u_{\nu} + P g_{\mu \nu} + \Sigma_{\mu \nu} ~,
\end{equation}
where $\rho$ and $P$ are the energy density and the pressure, respectively; $u_{\mu}$ is the four-velocity of the observer; $\Sigma_{\mu\nu}$ is the anisotropic stress subject to the conditions $\Sigma_{00}=\Sigma_{0i}=0$, $\Sigma_{ij}=\Sigma_{ji}$ and $\delta^{ij}\Sigma_{ij}=0$. For simplicity, we will set the anisotropic stress to zero and leave the study of the $\Sigma_{ij} \neq 0$ case to future work. From \eqref{eq:Tmunugeneral} we get
\begin{equation}
	\label{eq:Tij2}
	T_{ij}^{(2)} = ( \rho^{(0)} + P^{(0)} ) u_i^{(1)} u_j^{(1)}
	+ P^{(2)} \gamma_{ij}^{(0)} ~.
\end{equation}
The four-velocity by definition is normalized according to $u_{\mu} u_{\nu} g^{\mu \nu} = -1$. The zeroth-order four-velocity is
\begin{equation}
	u_0^{(0)} = -a ~,~ u^{(0)}_i = 0 ~.
\end{equation}
Along with the definition $u_{\mu} = g_{\mu \nu}u^{\nu}$, one can work out the four-velocity at any order perturbatively. We present some useful result here 
\begin{equation}
	u_i^{(1)} = 0 ~,~ u^{(1)i} = -\frac{1}{a} \partial^i \beta ~,
\end{equation}
where we used $\alpha = \zeta^{\prime}/\mathcal{H}$. The rest of the quantities are to be determined by the perturbed Einstein equations $G_{\mu \nu}^{(n)} = T_{\mu \nu}^{(n)}$. For instance,
\begin{equation}
	\rho^{(0)} = - T_{0}^{(0)0} = - G_{0}^{(0)0} = 3 \frac{\mathcal{H}^2}{a^2} ~,
\end{equation}
\begin{align}
	P^{(2)} &= \frac{1}{3} T_{i}^{(2)i} - \frac{1}{3} ( \rho^{(0)} + P^{(0)} ) u_i^{(1)} u^{(1)i} 
	\nn\\&= \frac{1}{3} G_{i}^{(2)i} - \frac{1}{3} ( \rho^{(0)} + P^{(0)} ) u_i^{(1)} u^{(1)i} ~.
\end{align}

We have shown the generic formalism to compute the Einstein equations order by order. Fortunately for our purpose, we do not need many of the above expressions. To see this, we define a projection vector to extract the transverse vector modes ~\cite{Chang:2022dhh}:
\begin{equation}
	\mathcal{V}_i^{kl} \equiv \frac{1}{\nabla^2} \partial^l \left( \delta_i^k - \frac{\partial^k \partial_i}{\nabla^2}  \right) ~,
\end{equation} 
and it is helpful to list the following relationships:
\begin{equation}
	\begin{aligned}
	&\mathcal{V}_i^{kl} \delta_{kl} = 0 ~,
	\quad
	\mathcal{V}_i^{kl} \partial_k \partial_l \Phi = 0 ~,
	\\&
	\mathcal{V}_i^{kl} \partial_k G_l = 0 ~,
	\quad
	\mathcal{V}_i^{kl} \partial_l G_k = G_i ~.
	\end{aligned}
\end{equation}
We immediately have
\begin{equation}
	\mathcal{V}_l^{ij} \left[ \frac{1}{2} \partial_i G_j^{\prime} + \mathcal{H} \partial_i G_j + ( i \longleftrightarrow j) \right] = \frac{1}{2}G_l^{\prime} + \mathcal{H} G_l ~.
\end{equation}
Additionally, the projection of $\delta_{ij} \mathcal{P}$ and $T_{ij}^{(2)}$ vanishes, which can be easily checked. The latter results from the fact of uniform gauge $\delta \phi = 0$. The projected Einstein equation simplifies to 
\begin{align}
	G_l^{\prime} + 2G_l = \mathcal{V}_l^{ij} S_{ij} ~,
\end{align}
where we define the source term to be 
\begin{align}
	S_{ij} = \frac{1}{2} (\partial_i \alpha \partial_j \alpha - \partial_i \zeta \partial_j \zeta) -& ( \partial_i \beta^{\prime} + \partial_i \alpha + 2 \mathcal{H} \partial_i \beta  ) \partial_j \zeta 
	\nn\\&+ ( i \longleftrightarrow j) ~.
\end{align}

As a final step, we move to the Fourier space. To describe the vector fluctuations, we choose a pair of polarization vectors $\{e^+(\hat{k}), e^{\times}(\hat{k}) \}$ , which are orthogonal to each other and $\vec{k}$, satisfying:
\begin{equation}
	\begin{aligned}
		&e_i^{\lambda}(\hat{k}) e^{\sigma,i}(\hat{k}) = \delta^{\lambda\sigma} ~,~ e_i^{\lambda}(\hat{k}) k^i = 0 ~,
		\\&
		\sum_{\lambda} e^{\lambda,i}(\hat{k}) e^{\lambda,j}(\hat{k})
		= \delta^{ij} - \frac{k^i k^j}{k^2} ~.
	\end{aligned}
\end{equation}
The vector perturbation becomes
\begin{equation}
	G_i(\tau,\vec{x}) = \sum_{\lambda} \int \frac{\ddd^3 \vec{k}} {(2\pi)^3} e^{i \vec{k} \cdot \vec{x}} G^\lambda(\tau,\vec{k}) e_i^{\lambda}(\hat{k}) ~,
\end{equation}
and we have 
\begin{align}
	\mathcal{V}_i^{ab} S_{ab}(\tau,\vec{x})
	&=
	\int \frac{\ddd^3 \vec{k}} {(2\pi)^3} \frac{i k^b}{k^2} \left( \delta_i^a - \frac{k^a k_i}{k^2} \right) e^{i \vec{k} \cdot \vec{x}} S_{ab} (\tau,\vec{k}) ~,
	\nn\\&= 
	\sum_{\lambda} \int \frac{\ddd^3 \vec{k}} {(2\pi)^3} \frac{i k^b}{k^2}
	e_i^{\lambda}(\hat{k}) e^{\lambda,a}(\hat{k}) e^{i \vec{k} \cdot \vec{x}} S_{ab} (\tau,\vec{k}) ~,
\end{align}
where $S_{ab} (\tau,\vec{k})$ is the Fourier transform of $S_{ab}(\tau,\vec{x})$. Hence the equation for the vector mode becomes
\begin{equation} 
	\label{eq:vectordynamicaleq_app}
	G^{\lambda \prime}(\tau,\vec{k}) + 2 \mathcal{H} G^{\lambda}(\tau,\vec{k}) = S^{\lambda}(\tau,\vec{k}) ~,
\end{equation}
where
\begin{equation}
	S^{\lambda} (\tau,\vec{k})
	= \frac{i k^a}{k^2} e^{\lambda,b}(\hat{k}) S_{ba} (\tau,\vec{k}) ~.
\end{equation}

In the end, we shall discuss the gauge of vector fluctuations. In the uniform field gauge, the SIVPs are solely determined by the antisymmetric part of $G_{ij}$. Adopting the gauge condition $G_i = 0$, we calculate
\begin{equation}
\begin{aligned}
	G_{ij}^{(2)} &= \frac{1}{2} \partial_i F_j^{\prime \prime} + \mathcal{H} \partial_i F_j^{\prime} - \frac{1}{2} \partial_i \alpha \partial_j \alpha + \frac{1}{2} \partial_i \zeta \partial_j \zeta 
	\\&+ ( \partial_i \beta^{\prime} + \partial_i \alpha + 2 \mathcal{H} \partial_i \beta  ) \partial_j \zeta + \partial_i \beta \partial_j \zeta^{\prime} \\
	& + \alpha \partial_i \partial_j ( \beta^{\prime} +  2 \mathcal{H} \partial_i \beta) + \frac{1}{2} (\partial^2 \beta + \alpha^{\prime} - \zeta^{\prime}) \partial_i \partial_j \beta 
	\\&- \frac{1}{2} \partial^a \beta \partial_a \partial_i \partial_j \beta  + ( i \longleftrightarrow j)  ~.
\end{aligned}
\end{equation}
Thus the gauge condition $F_i = 0$ and $G_i = 0$ gives the same result upon the identification $G_i = F_i^{\prime}$ of the two results. This should not be a surprise, since a gauge-invariant vector perturbation can be defined as $\sigma^i = G_i + F_i^{\prime}$ \cite{Durrer:1993tti}, and adopting either $F_i = 0$ or $G_i = 0$ will result in the same quantity.

Now we evaluate the Fourier transformation of $S_{ab}$. We take the term $\partial_a \zeta \partial_b\zeta$ as an example:
\begin{align}
	&{\rm FT}  (\partial_a \zeta (\tau) \partial_b \zeta (\tau)) 
	\nonumber\\& = -\int d^3x e^{-i\vec{k} \cdot \vec{x}} \int \frac{d^3p}{(2\pi)^3} \frac{d^3q}{(2\pi)^3} e^{i (\vec{p}+\vec{q}) \cdot \vec{x}} p_a q_b \zeta_{\vec{p}} (\tau) \zeta_{\vec{q}} (\tau) 
	\nn\\& = \int \frac{d^3p}{(2\pi)^3} p_a (p_b - k_b) \zeta_{\vec{p}} (\tau) \zeta_{\vec{k} - \vec{p}} (\tau) ~.
\end{align}
Thus, the projection procedure will result in the following structure:
\begin{equation}
	\frac{i k^a}{k^2} e^{\lambda,b}(\hat{k}) [p_a (p_b - k_b) + (a \leftrightarrow b)] = i e^{\lambda,b}(\hat{k}) p_b \left( 2\frac{k_ap^a}{k^2} - 1 \right) ~.
\end{equation}
The integration of $\int d^3 p p_b \zeta_{\vec{p}} (\tau) \zeta_{\vec{k} - \vec{p}} (\tau)$ will always give a vector parallel to $\vec{k}$, and vanishes after summation with $e^{\lambda,b}$, so we can write the above expression as
\begin{align}
	&\int d^3 p \frac{i k^a}{k^2} e^{\lambda,b}(\hat{k}) [p_a (p_b - k_b) + (a \leftrightarrow b)] \zeta_{\vec{p}} (\tau) \zeta_{\vec{k} - \vec{p}} (\tau) \nn\\&\to 
	\int d^3 p \frac{2i}{k^2} e^{\lambda}(\vec{k},\vec{p}) \zeta_{\vec{p}} (\tau) \zeta_{\vec{k} - \vec{p}} (\tau)  ~,
\end{align}
with
\begin{equation}
	e^{\lambda}(\vec{k},\vec{p}) \equiv k^ae^{\lambda,b}(\hat{k})p_bp_a ~.
\end{equation}
In the end
\begin{align}
	&\frac{i k^a}{k^2} e^{\lambda,b}(\hat{k}) {\rm FT} [\partial_a \zeta (\tau) \partial_b \zeta (\tau) + (a \leftrightarrow b)] 
	\nn\\&=  \int \frac{d^3p}{(2\pi)^3} \frac{2i}{k^2} e^{\lambda}(\vec{k},\vec{p}) \zeta_{\vec{p}} (\tau) \zeta_{\vec{k} - \vec{p}} (\tau) ~,
\end{align}
and
\begin{align}
	S^{\lambda} & \nonumber = i \int \frac{d^3p}{(2\pi)^3} \frac{e^{\lambda}(\vec{k},\vec{p})}{k^2} \left[ \alpha_{\vec{p}} \alpha_{\vec{k} - \vec{p}} - \zeta_{\vec{p}} \zeta_{\vec{k} - \vec{p}} \right. \\
	& \left. - 2 \zeta_{\vec{p}}^{\prime} \beta_{\vec{k} - \vec{p}}  - 2 (\alpha_{\vec{p}} + \beta_{\vec{p}}^{\prime} + 2 \mathcal{H} \beta_{\vec{p}}) \zeta_{\vec{k} - \vec{p}} \right] ~.
\end{align}

We may also write the constraint equation in Fourier space to simplify the above expression:
\begin{equation}
	\alpha_{\vec{k}} = \frac{\zeta_{\vec{k}}^{\prime}}{\mathcal{H}}  ~,~ \beta_{\vec{k}} = - \frac{\zeta_{\vec{k}}}{\mathcal{H}} - \frac{3\zeta_{\vec{k}}^{\prime}}{2c_s^2k^2}  ~;~ \beta_{\vec{k}}^{\prime} = - \alpha_{\vec{k}} - \frac{1}{2} \zeta_{\vec{k}} - \frac{3\zeta_{\vec{k}}^{\prime \prime}}{2c_s^2k^2} ~.
\end{equation}
Specifically,
\begin{equation}
	\alpha_{\vec{p}} + \beta_{\vec{p}}^{\prime} + 2 \mathcal{H} \beta_{\vec{p}} = - \frac{3}{2c_s^2p^2} \left( \zeta_{\vec{p}}^{\prime \prime} + 2 \mathcal{H} \zeta_{\vec{p}} + p^2 c_s^2 \zeta_{\vec{p}} \right) - \zeta_{\vec{p}} = -\zeta_{\vec{p}} ~,
\end{equation}
where the dynamical equation of $\zeta$,
\begin{equation}
	\zeta^{\prime \prime} + 2\mathcal{H} \zeta - c_s^2\partial^2 \zeta = 0 ~\to~ \zeta_{\vec{p}}^{\prime \prime} + 2 \mathcal{H} \zeta_{\vec{p}} + p^2 c_s^2 \zeta_{\vec{p}} = 0 ~,
\end{equation}
is applied and we are left with
\begin{align}
	S^{\lambda}= i \int \frac{d^3p}{(2\pi)^3} \frac{e^{\lambda}(\vec{k},\vec{p})}{k^2} \left( \alpha_{\vec{p}} \alpha_{\vec{k} - \vec{p}} + \zeta_{\vec{p}} \zeta_{\vec{k} - \vec{p}} - 2 \zeta_{\vec{p}}^{\prime} \beta_{\vec{k} - \vec{p}} \right)   ~.
\end{align}

\subsection{Generic form of the two-point correlation function}
Now we will temporarily write the above source term in the schematic form,
\begin{equation}
	\label{eq:Slambdageneral}
	S^{\lambda} \simeq \frac{i}{k^2} \int \frac{d^3p} {(2\pi)^3} e^{\lambda}(\vec{k},\vec{p}) f(k,p,\tau) \zeta_{\vec{p}} (\tau_0) \zeta_{\vec{k} - \vec{p}} (\tau_0)  ~,
\end{equation}
and evaluate the two-point correlation function of 
$G^{\lambda}$, defined as
\begin{equation}
	\langle G^{\lambda} (\vec{k}) G^{s} (\vec{p}) \rangle \equiv  (2\pi)^3 \delta(\vec{k}+ \vec{p}) \delta_{\lambda s}  \frac{2\pi^2}{k^3} \mathcal{P}_G (\tau,\vec{k}) ~,
\end{equation} 
in terms of \eqref{eq:Slambdageneral}. We will discuss how to determine the form of $f(k,p,\tau)$ in the following section.

The general solution of $G_i$ is given by
\begin{equation}
	G^{\lambda}(\tau,\vec{k}) = \frac{1}{a(\tau)^2} \int^{\tau} a(\tilde{\tau})^2 S^{\lambda}(\tilde{\tau},\vec{k}) \ddd\tilde{\tau} ~.
\end{equation}
Applying the bouncing background and specifying the integration range, we have
\begin{equation}
	G^{\lambda}(\tau,\vec{k}) = \int_{\tau_{\textrm{ini}}}^{\tau} d\tilde{\tau} \left( \frac{\tilde{\tau}}{\tau} \right)^4 S^{\lambda}(\tilde{\tau},\vec{k}) ~,
\end{equation}
where $\tau_{\rm ini}$ is the initial conformal time of the contraction phase. The two-point correlation function of $G^{\lambda}$ becomes
\begin{widetext}
\begin{align}
	& \quad \langle G^{\lambda}(\tau,\vec{k}) G^{s}(\tau,\vec{k}^{\prime}) \rangle \nonumber = \int_{\tau_{\rm ini}}^{\tau} d\tilde{\tau}_1 \int_{\tau_{\rm ini}^{\prime}}^{\tau} d\tilde{\tau}_2 \left( \frac{\tilde{\tau}_1}{\tau} \right)^4 \left( \frac{\tilde{\tau}_2}{\tau} \right)^4   \langle S^{\lambda}(\tilde{\tau}_1,\vec{k}) S^{s}(\tilde{\tau}_2,\vec{k}^{\prime}) \rangle \\
	& \nonumber = - \frac{1}{k^2k^{\prime 2}}\int_{\tau_{\rm ini}}^{\tau} d\tilde{\tau}_1 \int_{\tau_{\rm ini}^{\prime}}^{\tau} d\tilde{\tau}_2 \left( \frac{\tilde{\tau}_1}{\tau} \right)^4 \left( \frac{\tilde{\tau}_2}{\tau} \right)^4 \int \frac{d^3p}{(2\pi)^3} \int \frac{d^3q}{(2\pi)^3} e^{\lambda}(\vec{k},\vec{p}) e^{s}(\vec{k}^{\prime},\vec{q}) \\
	& \times f^{\ast}(\vec{p}, \vec{k}, \tilde{\tau}_1)
	f(\vec{q}, \vec{k}^{\prime}, \tilde{\tau}_2) \langle \zeta_{\vec{p}}(\tau_0) \zeta_{\vec{k} - \vec{p}} (\tau_0)\zeta_{\vec{q}} (\tau_0) \zeta_{\vec{k}^{\prime} - \vec{q}} (\tau_0) \rangle ~. 
\end{align}
\end{widetext}

Assuming a Gaussian distribution of curvature fluctuation and with the help of the definition of the scalar power spectrum, the contraction of the four-point correlator is decomposed as
\begin{align}
	& \quad \langle \zeta_{\vec{p}} \zeta_{\vec{k} - \vec{p}} \zeta_{\vec{q}} \zeta_{\vec{k}^{\prime} - \vec{q}}\rangle (\tau_0) \nonumber = \langle \zeta_{\vec{p}}\zeta_{\vec{q}} \rangle (\tau_0) \langle \zeta_{\vec{k} - \vec{p}}\zeta_{\vec{k}' - \vec{q}}\rangle (\tau_0)
	+ \langle \zeta_{\vec{p}}\zeta_{\vec{k}' - \vec{q}} \rangle (\tau_0)\langle \zeta_{\vec{k} - \vec{p}} \zeta_{\vec{q}} \rangle (\tau_0) \\
	& = (2\pi)^6 \frac{2\pi^2}{p^3} \frac{2\pi^2}{|\vec{k} - \vec{p}|^3}  \delta (\vec{k} + \vec{k}^{\prime}) \left[ \delta(\vec{p} + \vec{q}) + \delta (\vec{q} + \vec{k} - \vec{p}) \right] \mathcal{P}_{\zeta}(\vec{p},\tau_0) \mathcal{P}_{\zeta} (\vec{k} - \vec{p},\tau_0) ~,
\end{align}
and the above formula becomes
\begin{widetext}
\begin{align}
	& \quad \langle G^{\lambda}(\tau,\vec{k}) G^{s}(\tau,\vec{k}^{\prime}) \rangle \nonumber  = 2 \frac{(2\pi^2)^2}{k^4}\int_{\tau_{\rm ini}}^{\tau} d\tilde{\tau}_1 \int_{\tau_{\rm ini}^{\prime}}^{\tau} d\tilde{\tau}_2 \left( \frac{\tilde{\tau}_1}{\tau} \right)^4 \left( \frac{\tilde{\tau}_2}{\tau} \right)^4 \int d^3p  \\
	& \times e^{\lambda}(\vec{k},\vec{p}) e^{s}(\vec{k},\vec{p}) f^{\ast}(\vec{p}, \vec{k}, \tilde{\tau}_1)
	f(\vec{p}, \vec{k}, \tilde{\tau}_2) \frac{\mathcal{P}_{\zeta}(\vec{p},\tau_0) \mathcal{P}_{\zeta}(\vec{k} - \vec{p},\tau_0)}{p^3 |\vec{k} - \vec{p}|^3} ~.
\end{align}
\end{widetext}
We adopt the coordinates of two polarization vectors and $\vec{k}$ as
\begin{align}
	&e^+(\hat{k}) = (1, 0, 0) ~,~ e^{\times}(\hat{k}) = (0, 1, 0) ~,
	\\&
	\vec{k} = (0, 0, k) ~,~  \vec{p} = p (\sin\theta\cos\psi, \sin\theta\sin\psi, \cos\theta) ~,
\end{align}
and the angular integration can be evaluated as
\begin{equation}
	\begin{aligned}
		\int d^3p e^+(\vec{k},\vec{p}) e^{+}(\vec{k},\vec{p}) 
		&= \int d^3p e^{\times}(\vec{k},\vec{p}) e^{\times}(\vec{k},\vec{p})
		\\& = k^2\pi \int_0^{\infty} p^6dp \int_0^{\pi} \sin^3 \theta \cos^2 \theta d\theta ~,
	\end{aligned}
\end{equation}
\begin{equation}
	\int d^3p e^{+}(\vec{k},\vec{p}) e^{\times}(\vec{k},\vec{p}) = 0~.
\end{equation}

With the introduction of auxiliary variables
\begin{align}
	&x \equiv \frac{|\vec{k} - \vec{p}|}{k} ~,~ y \equiv \frac{p}{k} ~
	\\&
	z \equiv k\tau < 0 ~,~ z_{\rm ini} \equiv k \tau_{\rm ini} < 0 ~,~ z_0 \equiv k\tau_0 < 0 ~,
\end{align}
one gets
\begin{equation}
	dp = kdy ~,~ \cos \theta = \frac{1+y^2 = x^2}{2y} ~,~ d\cos \theta = -\frac{x}{y}dx ~,
\end{equation}
and
\begin{widetext}
\begin{align}
	\langle G^{\lambda}(\tau,\vec{k}) G^{s}(\tau,\vec{k}') \rangle & \nonumber = \frac{8\pi^5}{k^3} \delta^{\lambda s} \delta(\vec{k} + \vec{k}')
	\int_{0}^{\infty} \ddd y \int_{|1 - y|}^{1+y} \ddd x \frac{y^2}{x^2} \left[ 1 - \left( \frac{1+y^2 - x^2}{2y} \right)^2 \right] \\
	& \times \left( \frac{1+y^2 - x^2}{2y} \right)^2 \mathcal{P}_{\zeta}(ky,\tau_0) \mathcal{P}_{\zeta}(kx,\tau_0) \left| \int_{z_{\rm ini}}^{z} d\tilde{z} \frac{\tilde{z}^4}{z^4} f(x,y,\tilde{z}) \right|^2 ~.
\end{align}
\end{widetext}

Following the convention in the study of SIGWs, we further define
\begin{equation}
	s = \frac{y-x}{\sqrt{2}} ~,~ t = \frac{y+x}{\sqrt{2}} ~, 
\end{equation}
and the total vector power spectrum is calculated as
\begin{equation}
\begin{aligned}
	\label{eq:PGbare}
	&\mathcal{P}_G(\tau, \vec{k}) = \int_{\frac{1}{\sqrt{2}}}^{\infty} \ddd t \int_{-\frac{1}{\sqrt{2}}}^{\frac{1}{\sqrt{2}}} \ddd s \frac{(1-2s^2)(2t^2-1)(2st+1)^2}{4(t^2-s^2)^2 } \\ 
	& \times \mathcal{P}_{\zeta}\left(\frac{k}{\sqrt{2}}(t-s),\tau_0 \right) \mathcal{P}_{\zeta} \left(\frac{k}{\sqrt{2}}(t+s), \tau_0 \right)  \left|  \mathcal{I}(s,t,z) \right|^2 ~,
\end{aligned}
\end{equation}
where 
\begin{equation}
	\mathcal{I}(s,t,z) = \int_{z_{\rm ini}}^{z} d\tilde{z} \frac{\tilde{z}^4}{z^4} f(s,t,\tilde{z}) ~,
\end{equation}
is the time integral. We are interested in the vector power spectrum at the end of the contraction phase, thus we will need to evaluate
\begin{equation}
	\label{eq:timeint}
	\mathcal{I}(s,t,z_0) = \int_{z_{\rm ini}}^{z_0} dz \frac{z^4}{z_0^4} f(s,t,z) ~.
\end{equation}

\subsection{Evaluating the vector power spectrum}
We remind the reader that in matter-bounce scenario, the quadratic action of curvature fluctuation in k-essence theory takes the following form:
\begin{equation}
	S_{\zeta}^{(2)} = \int d\tau d^3x \frac{3a^2}{2c_s^2} \left[ \zeta^{\prime 2} - c_s^2 (\partial_i \zeta)^2 \right] ~.
\end{equation}
In Fourier space the solution is
\begin{equation}
	\zeta_{\vec{k}} (\tau) = \frac{e^{-ik c_s \tau} c_s}{\sqrt{6c_sk}} \left( 1 - \frac{i}{c_s k \tau} \right) \left( \frac{\tau_0}{\tau} \right)^2 ~.
\end{equation}
One may use this expression to work out $\alpha_{\vec{k}}$ and $\beta_{\vec{k}}$, and then the expression of $f(k,p,\tau)$:
\begin{equation}
	f(k,p,\tau) \equiv \frac{\alpha_{\vec{p}} (\tau) \alpha_{\vec{k} - \vec{p}} (\tau) + \zeta_{\vec{p}} (\tau) \zeta_{\vec{k} - \vec{p}} (\tau) - 2 \zeta_{\vec{p}}^{\prime} (\tau) \beta_{\vec{k} - \vec{p}} (\tau)}{\zeta_{\vec{p}} (\tau_0) \zeta_{\vec{k} - \vec{p}} (\tau_0)} ~.
\end{equation}

Nevertheless, the general expressions are quite involved after the integration. Additionally, we may count in the contributions from sub-horizon modes, which could just be quantum zero-point energy, and it is not clear whether it makes sense to collect their contributions to SIVPs. Therefore, in the following, we will use the following expression:
\begin{align}
	\label{eq:zetakreg}
	&\zeta_k (\tau) \simeq\frac{e^{-ik c_s \tau} c_s}{\sqrt{6c_sk}} \left(  - \frac{i}{c_s k \tau} \right) \left( \frac{\tau_0}{\tau} \right)^2  ~,
	\\&
	 \zeta_k^{\prime} \simeq -\left(\frac{3}{\tau} + ikc_s \right)\zeta_k = - \left(\frac{3}{2} \mathcal{H} + ikc_s \right) \zeta_k  > 0 ~,
\end{align}
corresponding to the curvature fluctuation with the removal of the leading counter term in adiabatic regularization formalism \cite{Zeldovich:1971mw,Parker:1974qw}. It is easy to check that the canonically normalized fluctuation $v_k \equiv \zeta_k/z_s$ vanishes in the far past $\tau \to \infty$, namely the zero-point energy from the oscillatory solution $v_k^{\prime \prime} - c_s^2k^2 = 0$ is removed. Although we have no intention to judge  whether such contributions are appropriate to be taken into account, the usage of \eqref{eq:zetakreg} would not lead to overestimation of the SIVPs, since it simply erases the contributions from the case $k|\tau| \gg 1$. Apparently, Eq. \eqref{eq:zetakreg} is not the rigid way to perform SIVPs regularization. However, the regularization problem is quite involved even in inflation (see, e.g., Ref. \cite{Ye:2022tgs}), and rarely studied in bouncing cosmology. Also, when one tries to include more counter terms for either the curvature or the vector fluctuations, the general expressions become highly complicated, which further leads to an artificial divergence of the momentum integral. We will work with \eqref{eq:zetakreg} in this paper, and it is expected that our work may motivate people to think more deeply about the regularization issue of bouncing cosmology, as its significance is highlighted in our paper. 

Now, the constraint equations simplify to
\begin{widetext}
\begin{equation}
	\label{eq:constraintexp}
	\alpha_k = \frac{\zeta_k^{\prime}}{\mathcal{H}} \simeq - \left( \frac{3}{2} + \frac{ikc_s}{\mathcal{H}} \right) \zeta_k ~,~ \beta_k = - \frac{\zeta_k}{\mathcal{H}} + \frac{3\zeta_k^{\prime}}{2c_s^2k^2} = - \frac{\zeta_k}{\mathcal{H}} - \frac{3 \mathcal{H} }{2c_s^2k^2} \left( \frac{3}{2} + \frac{ikc_s}{\mathcal{H}} \right) \zeta_k ~,
\end{equation}
and
\begin{align}
	& f(s,t,z) \simeq \left[ \frac{1}{4} \left( 1 + ic_s z (x+y) - c_s^2xyz^2 \right) - \frac{(x^2+y^2) ( 3 + ic_s xz) ( 3 + ic_s yz)}{2c_s^2x^2y^2z^2} \right] \left( \frac{z_0}{z} \right)^6 e^{-i\sqrt{2}t c_s z} 
	\nn\\
	& = \frac{e^{-i\sqrt{2}t c_s z}}{4} \left( \frac{z_0}{z} \right)^6 \left\{ 1 + \sqrt{2}i c_s tz + \frac{c_s^2z^2}{2}(t^2-s^2) - \frac{4(t^2+s^2)}{c_s^2z^2(t^2-s^2)^2}[3\sqrt{2}i+c_sz(t-s)] [3\sqrt{2}i+c_sz(t+s)]\right\} ~,
\end{align}
\end{widetext}
up to an overall phase factor with the following approximation
\begin{equation}
	\frac{ \zeta_{\vec{k}} (\tau) }{\zeta_{\vec{k}} (\tau_0)} \simeq \frac{\tau_0^3}{\tau^3} e^{-icsk (\tau - \tau_0)} = \left( \frac{z_0}{z} \right)^3 e^{-ic_s z} ~,
\end{equation}
and the fact that
\begin{equation}
	\frac{k}{\mathcal{H}} = k\tau/2=z/2 ~;~ \frac{p}{\mathcal{H}} = yz/2 ~,~ \frac{|\vec{k} - \vec{p}|}{\mathcal{H}} = xz/2 ~.
\end{equation}
We also write the above result in a symmetric form. The time integral now can be integrated out analytically and we define its primitive function to be
\begin{widetext}
\begin{align}
	& \quad  \nonumber \mathcal{F}(\tilde{z}) \equiv \int^{\tilde{z}} dz \frac{z^4}{z_0^4} f(s,t,z) \\
	& = \frac{z_0^2}{4} e^{-\sqrt{2}itc_s \tilde{z}} \left[ \frac{ic_s (s^2-t^2)}{2\sqrt{2}} + \frac{24(s^2+t^2)}{c_s^2 (s^2-t^2)^2 z^3} + \frac{3s^2 + 5t^2}{(s^2-t^2)z}  \right] + \frac{\sqrt{2}ic_st(t^2+s^2)}{(s^2-t^2)} z_0^2 {\rm Ei} (-i\sqrt{2}c_s t z) ~.
\end{align}
\end{widetext}
where ${\rm Ei}(x)$ is the exponential integral function.

\subsection{SIVPs in the constant curvature gauge}

The constant curvature gauge is defined by $\delta \mathcal{R} = 0$, where $\delta \mathcal{R}$ is the perturbed curvature scalar on the hypersurface. It is proved in \cite{Vitenti:2011yc} that this gauge is robust in the study of bouncing cosmology, thus we will evaluate SIVPs in this gauge and compare the result with that above. 

We refer the reader to Ref. \cite{Vitenti:2011yc} for more details on the constant curvature gauge and simply conclude their result here. The perturbed metric with only scalar fluctuations is
\begin{equation}
	ds^2 = a^2 \left[ -e^{-2\phi} d\tau^2 - \partial_i \mathcal{B} dx^id\tau + e^{2\psi}e^{ -2 \partial_i \partial_j \mathcal{E}} dx^idx^j \right] ~,
\end{equation}
and in the constant curvature gauge one utilizes $\psi = \mathcal{B} = 0$. Accordingly
\begin{equation}
	\phi = \frac{3}{2} \zeta ~,~ \mathcal{E} = \int d\tau \frac{\Phi}{\mathcal{H}} ~,
\end{equation}
with $\Phi$ the Bardeen potential and we have used $p = 0$ in matter-contraction phase so that $\beta$ defined in \cite{Vitenti:2011yc} simplifies to $3\mathcal{H}^2/2$. 

The Bardeen potential $\Phi$ is related to the curvature fluctuation via~\cite{Bardeen:1980kt,Brandenberger:1983tg}
\begin{equation}
	\zeta = \Phi + \frac{\mathcal{H}}{\mathcal{H}^2 - \mathcal{H}^{\prime}} ( \Phi^{\prime} + \mathcal{H} \Phi ) ~.
\end{equation}
In matter-contraction phase it reduces to 
\begin{equation}
	\zeta = \frac{5}{3} \Phi + \frac{\tau}{3} \Phi^{\prime} ~.
\end{equation}
Using \eqref{eq:zetak} we get $\Phi$ in the Fourier domain to be
\begin{equation}
	\Phi_k = \sqrt{\frac{3c_s}{2k}} \left( \frac{\tau_0}{\tau} \right)^2 e^{-ic_sk\tau} \frac{-3i + 3c_sk\tau + i c_s^2 k^2 \tau^2}{(c_sk\tau)^3} ~,
\end{equation}
where the integration constant can be fixed by using the dynamical equation of $\zeta$. 

We can integrate out $\mathcal{E}$ analytically, whose primitive function is
\begin{align}
	\int d\tau \frac{\Phi}{\mathcal{H}} & = -\sqrt{\frac{3c_s}{8k}} \left( \frac{\tau_0}{\tau} \right)^2  \frac{e^{-ic_s k\tau}}{(c_s k)^2} \left( 1 - \frac{i}{c_s k \tau} \right) ~,
\end{align}
and we are free to choose the integration range as long as $\mathcal{E}$ remains small in the domain we considered. There is a particularly convenient choice by setting $\mathcal{E} = 0$ to be proportional to $\zeta$:
\begin{equation}
	\mathcal{E}_k = \frac{3}{2(c_sk)^2} \zeta_k ~,~ \partial^2 \mathcal{E} (\vec{x}) = -\frac{3\zeta (\vec{x})}{2c_s^2} ~,
\end{equation}
in Fourier space and normal space, respectively.

Now we will introduce the vector fluctuation. This time it is more convenient to choose $G_i = 0$, and the metric simplifies to
\begin{equation}
	ds^2 = a^2 \left[ -e^{-3\zeta} d\tau^2 + e^{ -2 \partial_i \partial_j \mathcal{E} + F_{ij}} dx^idx^j \right] ~.
\end{equation}
Of course, we can proceed with the gauge $F_i = 0$ and find identical results, similar to what we did in the case of the uniform field gauge. For the reader's convenience, we write down the spatial component of the metric up to second order
\begin{equation}
	e^{ -2 \partial_i \partial_j \mathcal{E} + F_{ij}} = \delta_{ij} - 2 \partial_i \partial_j \mathcal{E} + 2 \partial_i \partial_k  \mathcal{E}  \partial^k \partial_j \mathcal{E} + F_{ij} ~,
\end{equation}
and we remind the reader that $F_{ij}$ itself is second order. 

The Einstein tensor at second order is quite lengthy. For our purpose, we only need the parts that are nonvanishing after the projection. Using the identity 
\begin{equation}
	\mathcal{V}_i^{kl} \delta_{kl} = 0 ~,~ \mathcal{V}_i^{kl} \partial_k \partial_l \Phi = 0 ~,~ \mathcal{V}_i^{kl} \partial_k G_l = 0 ~,~
	\mathcal{V}_i^{kl} \partial_l G_k = G_i ~,
\end{equation}
we summarize 
\begin{widetext}
\begin{align}
	\mathcal{V}_i^{kl} G_{kl}^{(2)} & = \frac{1}{2} F_i^{\prime \prime} + \mathcal{H} F_i^{\prime} - \frac{9}{4} \mathcal{V}_i^{kl} \partial_k \zeta \partial_l \zeta 
	+ \mathcal{V}_i^{kl} [\partial^a \partial_l (\partial^b \partial_a \mathcal{E} \partial_b \partial_k \mathcal{E}) + (l \leftrightarrow k) ] 
	\nn\\
	&+ \mathcal{V}_i^{kl} \Big[ \partial_a \partial_k (\mathcal{E}^{\prime \prime} + 2\mathcal{H} \mathcal{E}^{\prime } - \partial^2 \mathcal{E}) \partial_a \partial_l \mathcal{E} 
	+ \partial_a \partial_k \mathcal{E} \partial_a \partial_l (\mathcal{E}^{\prime \prime} + 2\mathcal{H} \mathcal{E}^{\prime } - \partial^2 \mathcal{E}) \Big] ~.
\end{align}
\end{widetext}
We have kept the term $\mathcal{V}_i^{kl} \partial^a \partial_k (\partial^b \partial_a \mathcal{E} \partial_b \partial_l \mathcal{E})$ which vanishes due to the projection for later convenience. The dynamical equation of $\zeta$ leads to
\begin{equation}
	\zeta^{\prime \prime} + 2\mathcal{H} \zeta - c_s^2\partial^2 \zeta = 0 ~\to~ \mathcal{E}^{\prime \prime} + 2\mathcal{H} \mathcal{E}^{\prime } - c_s^2 \partial^2 \mathcal{E} = 0 ~,
\end{equation}
which further simplifies the result as
\begin{align}
	\mathcal{V}_i^{kl} G_{kl}^{(2)} & = \frac{1}{2} F_i^{\prime \prime} + \mathcal{H} F_i^{\prime} - \frac{9}{4} \mathcal{V}_i^{kl} \partial_k \zeta \partial_l \zeta 
	\nn\\&+ \mathcal{V}_i^{kl} [ \partial^a \partial_l (\partial^b \partial_a \mathcal{E} \partial_b \partial_k \mathcal{E}) + (l \leftrightarrow k) ] 
	\nn\\& + (c_s^2 - 1) \mathcal{V}_i^{kl} \left[ \partial_a \partial_k (\partial^2 \mathcal{E}) \partial_a \partial_l \mathcal{E} + \partial_a \partial_k \mathcal{E} \partial_a \partial_l (\partial^2 \mathcal{E})  \right] ~.
\end{align}
Note that
\begin{align}
	\partial^a \partial_l (\partial^b \partial_a \mathcal{E} \partial_b \partial_k \mathcal{E}) 
	&= (\partial^a \partial_l \partial^b \partial_a \mathcal{E}) \partial_b \partial_k \mathcal{E} + \partial^b \partial_a \mathcal{E} (\partial^a \partial_l \partial_b \partial_k \mathcal{E} ) 
	\nn\\&= \partial_l \partial^b \partial^2 \mathcal{E} \partial_b \partial_k \mathcal{E} + \partial^b \partial_a \mathcal{E} (\partial^a \partial_b \partial_l \partial_k \mathcal{E} ) ~,
\end{align}
so
\begin{align}
	&[ \partial^a \partial_l (\partial^b \partial_a \mathcal{E} \partial_b \partial_k \mathcal{E}) - \partial_a \partial_k \mathcal{E} \partial_a \partial_l (\partial^2 \mathcal{E}) + (l \leftrightarrow k) ] 
	\nn\\&= 2 \partial^b \partial_a \mathcal{E} (\partial^a \partial_b \partial_l \partial_k \mathcal{E} ) ~,
\end{align}
which vanishes after the projection due to the $\partial_l \partial_k \mathcal{E}$ term. We finally arrive at
\begin{widetext}
\begin{equation}
	\mathcal{V}_i^{kl} G_{kl}^{(2)} \to \frac{1}{2} F_i^{\prime \prime} + \mathcal{H} F_i^{\prime} -  \mathcal{V}_i^{kl} \left\{ \frac{9}{4} \partial_k \zeta \partial_l \zeta - c_s^2 \left[ \partial_a \partial_k (\partial^2 \mathcal{E}) \partial_a \partial_l \mathcal{E} + \partial_a \partial_k \mathcal{E} \partial_a \partial_l (\partial^2 \mathcal{E})  \right] \right\}
\end{equation}
\end{widetext}

We will also need the matter sector. Recall that
\begin{equation}
	T_{ij}^{(2)} = ( \rho^{(0)} + P^{(0)} ) u_i^{(1)} u_j^{(1)}
	+ P^{(2)} h_{ij}^{(0)} ~,
\end{equation}
and the $P^{(2)} h_{ij}^{(0)}$ term vanishes after the projection since $h_{ij}^{(0)} \propto \delta_{ij}$. We can also work out
\begin{equation}
	u_i^{(1)} = - \frac{a\partial_i \zeta}{\mathcal{H}} ~\to~ \mathcal{V}_i^{kl} T_{kl}^{(2)} = 3\mathcal{V}_i^{kl} \partial_i \zeta \partial_j \zeta ~. 
\end{equation}
The projected Einstein equation $\mathcal{V}_i^{kl} (G_{kl}^{(2)} - T_{ij}^{(2)})= 0$ thus simplifies to
\begin{equation}
	G^{\lambda \prime}(\tau,\vec{k}) + 2 \mathcal{H} G^{\lambda}(\tau,\vec{k}) = S^{\lambda}(\tau,\vec{k}) ~,
\end{equation}
Here we have changed back to $G_i$ using $F_i^{\prime} \to G_i$ discussed above to have similar form with \eqref{eq:vectordynamicaleq}, and in this case $S^{\lambda}(\tau,\vec{k})$ has the expression
\begin{equation}
	S^{\lambda} (\tau,\vec{k})
	= \frac{i k^a}{k^2} e^{\lambda,b}(\hat{k}) S_{ba} (\tau,\vec{k}) ~,
\end{equation}
\begin{widetext}
	\begin{equation}
		S_{ba} (\tau,\vec{k}) = {\rm FT} \left\{ \frac{21}{4} \partial_b \zeta \partial_a \zeta  - c_s^2 \left[ \partial_i \partial_b (\partial^2 \mathcal{E}) \partial_i \partial_a \mathcal{E} + \partial_i \partial_b \mathcal{E} \partial_i \partial_a (\partial^2 \mathcal{E}) \right] \right\} ~.
	\end{equation}
\end{widetext}

The Fourier transformation of $\partial_b \zeta \partial_a \zeta$ has already been evaluated. We illustrate the Fourier transformation the other terms
\begin{widetext}
\begin{align}
	& \quad {\rm FT} [ \partial_i \partial_b (\partial^2 \mathcal{E}) \partial_i \partial_a \mathcal{E} + \partial_i \partial_b \mathcal{E}
	\partial_i \partial_a (\partial^2 \mathcal{E}) ] 
	\nn\\
	& = -\int d^3x e^{-i\vec{k} \cdot \vec{x}} \int \frac{d^3p}{(2\pi)^3} \frac{d^3q}{(2\pi)^3} e^{i (\vec{p}+\vec{q}) \cdot \vec{x}} p_ip_b q_iq_a 
	(p^2+q^2) \mathcal{E}_{\vec{p}} (\tau) \mathcal{E}_{\vec{q}} (\tau) 
	\nn\\
	&= - \int \frac{d^3p}{(2\pi)^3} (p^2+q^2) p^i (k_i-p_i) p_b (k_a - p_a) \mathcal{E}_{\vec{p}} (\tau) \mathcal{E}_{|\vec{k} - \vec{p}|} (\tau) 
	\nn\\
	& = -\frac{9}{4c_s^4} \int \frac{d^3p}{(2\pi)^3} \frac{p^2 + |\vec{k} - \vec{p}|^2}{p^2|\vec{k} - \vec{p}|^2} p^i 
	(k_i-p_i) p_b (k_a - p_a) \zeta_{\vec{p}} (\tau) \zeta_{|\vec{k} - \vec{p}|} (\tau) ~.
\end{align}
\end{widetext}
Let us work out the projection:
\begin{align}
	&\frac{i k^a}{k^2} e^{\lambda,b}(\hat{k}) p^i (k_i-p_i) p_b (k_a - p_a) 
	\nn\\&= (\vec{k} \cdot \vec{p}) e^{\lambda,b}(\hat{k})p_b \frac{k^2+p^2}{k^2} - e^{\lambda,b}(\hat{k})p_b \frac{k^2p^2 + (\vec{k} \cdot \vec{p})^2}{k^2} ~.
\end{align}
We comment that the latter term will introduce an additional factor $\cos \theta$ or $(\cos \theta)^{-1}$ compared to the first term and will lead to the vanishing of angular integration. Therefore, in terms of the tensor
\begin{equation}
	e^{\lambda}(\vec{k},\vec{p}) \equiv k^ae^{\lambda,b}(\hat{k})p_bp_a = (\vec{k} \cdot \vec{p}) e^{\lambda,b}(\hat{k})p_b  ~,
\end{equation}
we get the nonvanishing terms
\begin{widetext}
\begin{equation}
	S^{\lambda} (\tau,\vec{k}) = \frac{i}{k^2} \int \frac{d^3p}{(2\pi)^3} e^{\lambda}(\vec{k},\vec{p}) \zeta_{\vec{p}} (\tau) \zeta_{|\vec{k} - \vec{p}|} (\tau) \left[ \frac{21}{4} + \frac{9(1+y^2)}{4c_s^2} \left( \frac{1}{x^2} + \frac{1}{y^2} \right) \right] ~.
\end{equation}
\end{widetext}
This enables us to directly read out the quantity
\begin{equation}
	f(\vec{p},\vec{k},\tau) = \frac{3}{4} \left[ 7 + \frac{6}{c_s^2} \frac{(t^2+s^2) (2+(s+t)^2)}{(t^2-s^2)^2} \right] \left( \frac{\tau_0}{\tau} \right)^6 e^{ic_s k(\tau_0 - \tau)} ~.
\end{equation}
The primitive function of the time integral is
\begin{widetext}
\begin{equation}
	\mathcal{F}(z) = - \frac{3z_0^2}{4z}e^{-i\sqrt{2}c_s z} \left[ 7 + \frac{6}{c_s^2} \frac{(t^2+s^2) (2+(s+t)^2)}{(t^2-s^2)^2} \right] \left[ 1 + i\sqrt{2}c_s t z e^{i\sqrt{2}c_s z}{\rm Ei}(-i\sqrt{2}c_s t z ) \right] ~.
\end{equation}
\end{widetext}

\begin{figure}[htbp]
	\centering
	\includegraphics[width=0.8\linewidth]{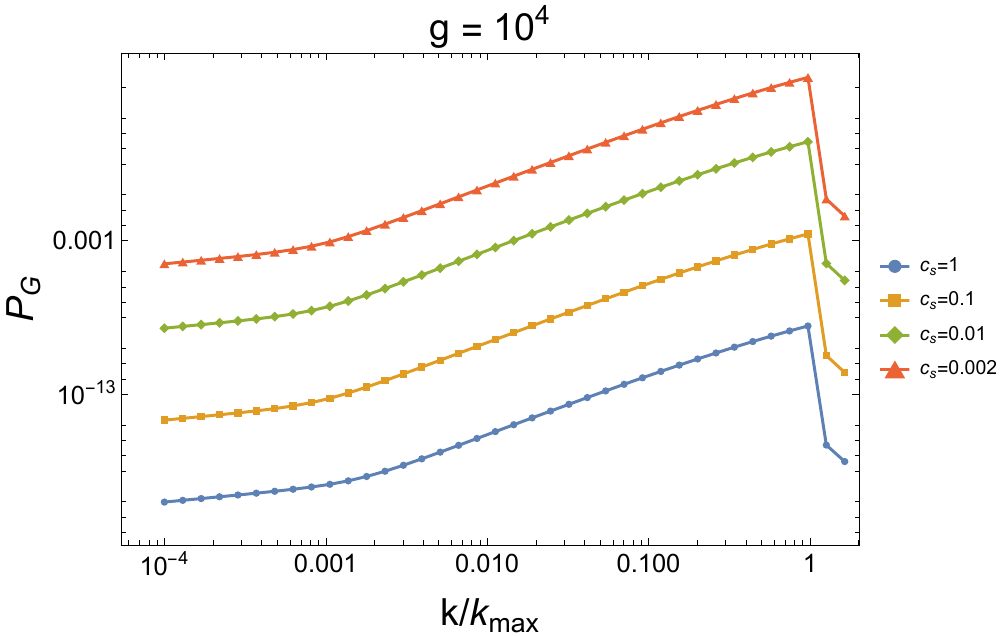} \\
	\includegraphics[width=0.8\linewidth]{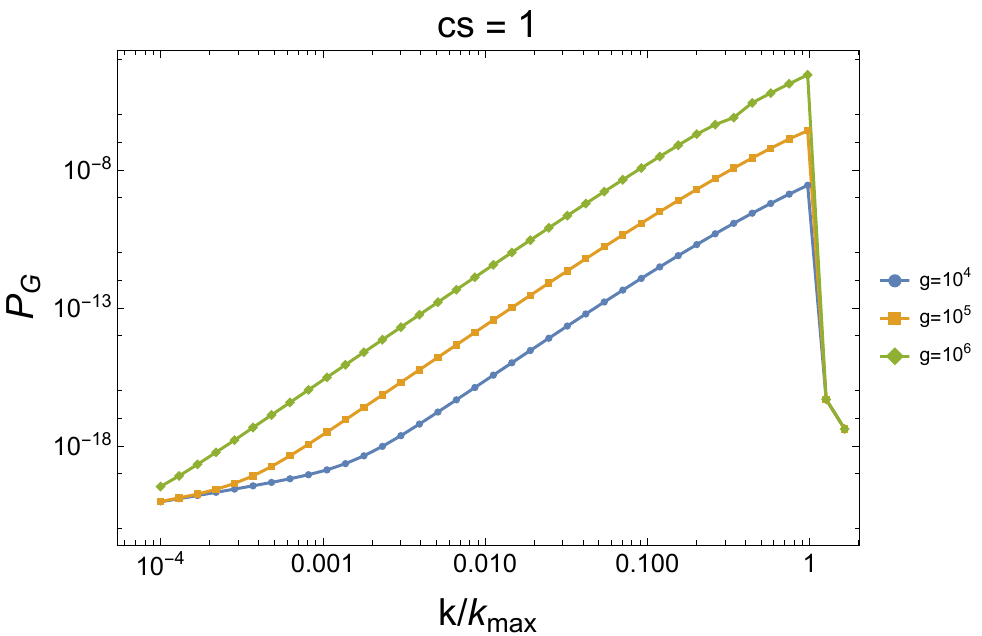}
	\caption{The vector power spectrum in the constant curvature gauge. The upper channel shows $\mathcal{P}_G$ with $g = 10^4$ fixed, and the lower channel shows $\mathcal{P}_G$ with $c_s = 1$ fixed.}
	\label{fig:PGccg}
\end{figure}

It is straightforward to compute SIVPs numerically using the above results. We depict the vector power spectrum in Fig. \ref{fig:PGccg}, which has a similar power law parametrization expect for the $k/k_{\rm max} \ll 1$ case when $g$ is relatively small: 
\begin{equation}
	\mathcal{P}_G(\tau_0,k) = \mathcal{A}_G \left( \frac{k}{k_{\rm max}} \right)^{n_G} ~,~ k < k_{\rm max} ~.
\end{equation}
The ratio of $\rho_V$ versus the background energy density value $\rho_{\rm bg}$ is also
\begin{equation}
	\delta_V(\tau_0) = \frac{\mathcal{A}_G}{6c_s^2 (n_G + 2)} ~.
\end{equation}

\begin{table}[htbp]
	\centering
	\begin{tabular}{|c|c|c|c|c|}
		\hline
		\multicolumn{2}{|c|}{Model parameters} & \multicolumn{3}{|c|}{Outcomes} \\
		\hline 
		$c_s$ & $g$ & $\mathcal{A}_G$ & $n_G$ & $\delta_V (\tau_0)$ \\ 
		\hline
		$2 \times 10^{-3}$ & $10^4$ & $3.6 \times 10^{7}$ & $3.6$ & $2.7 \times 10^{11}$ \\  
		\hline
		$10^{-2}$ & $10^4$ & $2.3 \times 10^{3}$ & $3.6$ & $6.8 \times 10^{5}$ \\  
		\hline
		$10^{-1}$ & $10^4$ & $2.3 \times 10^{-3}$ & $3.6$ & $6.8 \times 10^{-3}$ \\ 
		\hline
		$1$ & $10^4$ & $2.3 \times 10^{-9}$ & $3.6$ & $6.8 \times 10^{-11}$ \\
		\hline
		$1$ & $10^5$ & $2.3 \times 10^{-7}$ & $3.6$ & $6.8 \times 10^{-9}$ \\
		\hline
		$1$ & $10^6$ & $2.3 \times 10^{-5}$ & $3.6$ & $6.8 \times 10^{-7}$ \\
		\hline
	\end{tabular}
	\caption{Vector power spectra and the values of $\delta_V$ for various values of $c_s$ and $g$ in the constant curvature gauge.}
	\label{tab:PGCCG}
\end{table}	

Similarly, we also organize the values of $\mathcal{A}_G$ and $\delta_V (\tau_0)$ in Table \ref{tab:PGCCG}. Interestingly, the vector power spectrum is blue tilted and we have the scaling $\mathcal{A}_G \propto c_s^{-6} g^2$ and $\delta_V \propto A_s^2g^2c_s^{-8}$. It is quite straightforward to see that the energy density in the constant curvature gauge is much larger than that in the uniform field gauge considering $c_s < 0.018$ and we conclude that the SIVPs are also overproduced in this gauge.

\subsection{Comparison of the results}

Finally, let us comment on the difference of the above results. While one naturally expects SIVPs to be different from different gauge choices, it is also surprising that the two gauges yield such different results: in uniform field gauge the vector power spectrum is nearly scale invariant, whereas in constant curvature gauge it is highly blue. 

To study this problem, we note that both gauges indicate the same structure of SIVPs, and the only difference lies in the form of $f(k,p,\tau)$. Furthermore, as we have seen above, the main contribution to the momentum integral comes from the region $x \to 0$ or $y \to 0$. We write down the expression for $f$ in both gauges:
\begin{equation}
	f(k,p,\tau)_{\rm UFG} = -\frac{9}{2} \frac{(x^2+y^2)}{c_s^2 x^2y^2} \frac{1}{z^2} e^{-i\sqrt{2}t c_s z} \left( \frac{z_0}{z} \right)^6 + (...) ~,
\end{equation}
\begin{equation}
	f(k,p,\tau)_{\rm CCG} = \frac{9}{4} \frac{(x^2+y^2)}{c_s^2 x^2y^2} e^{-i\sqrt{2}t c_s z} \left( \frac{z_0}{z} \right)^6 + (...) ~,
\end{equation}
where $(...)$ stands for terms that contain less power of $x^{-1}$ or $y^{-1}$. 

Therefore, $f(k,p,\tau)_{\rm UFG}$ has an additional $z^2 = (k\tau)^2$ factor compared to $f(k,p,\tau)_{\rm CCG}$. We immediately see that the $n_G$ from the uniform field gauge shall receive an additional correction of 4 compared to that in the constant curvature gauge, as the time integral shall be squared before performing the momentum integral. In the numerical result, we get a difference of $3.6$ instead of $4$ between the two $n_G$, since the omitted terms also contribute to the power spectrum.

Additionally, we may write
\begin{align}
	&\int^{z_0} dz \frac{e^{-i\sqrt{2}t c_s z}}{z^n} 
	\nn\\&= (i\sqrt{2}t c_s)^{n-1} \int^{z_0} d(i\sqrt{2}t c_s z) \frac{e^{-i\sqrt{2}t c_s z}}{(i\sqrt{2}t c_s z)^n} 
	\nn\\&= (i\sqrt{2}t c_s)^{n-1} \int^{-2 \frac{k}{k_{\rm max}}} du \frac{e^{-iu}}{u^n} ~,
\end{align}
where we used
\begin{equation}
	c_s z_0 = c_s k \tau_0 = (c_s k_{\rm max} \tau_0) \frac{k}{k_{\rm max}} = -2 \frac{k}{k_{\rm max}} ~.
\end{equation}
We can immediately find that this $z^{-2}$ term in the uniform field gauge will lead to an additional $c_s^2$ in the time integral and thus an additional $c_s^4$ factor in the vector power spectrum, which agrees with our numerical result.

We may delve into the emergency of the $z^{-2}$ factor in the uniform field gauge. Recall that the $\frac{(x^2+y^2)}{c_s^2 x^2y^2} \frac{1}{z^2}$ term comes from the $\zeta_{\vec{p}}^{\prime} \beta_{\vec{k} - \vec{p}}$ term in the Einstein equation. In the superhorizon limit we approximately have
\begin{equation}
	\partial_a (\zeta_{\vec{p}}^{\prime}) \partial_b (\beta_{\vec{k} - \vec{p}}) \simeq - p_aq_b \frac{27}{8c_s^2} \frac{\mathcal{H}^2}{|\vec{k} - \vec{p}|^2} \zeta_{\vec{p}} \zeta_{\vec{k} - \vec{p}} ~.
\end{equation}
On the other hand, the $\frac{(x^2+y^2)}{c_s^2 x^2y^2} $ term in the constant curvature gauge originates from the $c_s^2\partial_i \partial_b (\partial^2 \mathcal{E}) \partial_i \partial_a \mathcal{E}$ term in Einstein equation, which can be written as
\begin{equation}
	{\rm FT} [c_s^2 \partial_i \partial_b (\partial^2 \mathcal{E}) \partial_i \partial_a \mathcal{E}] \to   -\frac{9}{4c_s^2} p_aq_b \frac{\vec{p} \cdot \vec{q}}{p^2} \zeta_{\vec{p}} \zeta_{\vec{k} - \vec{p}} ~.
\end{equation}
We see that these two terms differ from each other by a factor proportional to $\mathcal{H}^2/k^2$, which is exactly $z^{-2}$. Note that, the SIGW is generated at the horizon reentry event, and we have $\mathcal{H}(t_k)^2/k^2 = c_s^2$, which is merely a constant. In both situations, we are evaluating SIGW at different time slices, so this $\mathcal{H}^2/k^2$ factor will not lead to a structural difference in the time integral. In matter-contraction scenario, however, the curvature fluctuation grows on the superhorizon scale and we have to evaluate SIVPs at a specific time slice $\tau = \tau_0$ for all fluctuation modes. Therefore, the $\mathcal{H}^2/k^2$ factor will indeed introduce differences in the spectra index of the vector power spectrum.

From the above calculations, we see that the SIVPs are indeed highly dependent on the choice of gauges, at least in the matter-bounce scenario. Fortunately, we find the overproduction of SIVPs in both gauges in realistic matter-bounce models, and we can still claim that matter-bounce scenario can be constrained by SIVPs.

\subsection{Interpretation of our results in the two different approaches}

Let us first review the origin of gauge dependence in the case of SIGWs. From a theoretical perspective, scalars (S), vectors (V) and tensors (T) can be defined by their transformation properties. When taking different gauges, the split of space and time would lead to different definitions of SVTs. Thus, a gauge-invariant formulation of the SVTs does not help at all at the secondary level. One must define the SVTs in a particular gauge choice before defining gauge-invariant variables. At linear level the gauge independence of scalar, vector and tensors comes from the SVT decomposition. At the nonlinear level, ambiguities take the presence due to the mixture of those modes. For example, the tensor modes transform according to Eq. (7.6) in Ref. \cite{Domenech:2021ztg}, which is indeed gauge dependent. Additionally, the energy density of SIGWs, which is supposed to be an observable, is gauge dependent. 

To the best of our knowledge, the gauge issue of SIGWs remains to be fully addressed. Currently, one promising direction is to argue what is the gauge (or coordinate frame) that best describes the GW detection. According to \cite{DeLuca:2019ufz}, the best gauge is the so-called transverse-traceless (TT) gauge based on the analogy with asymptotically flat spacetimes. In cosmological setups, the closest gauge to the TT gauge in a cosmological background is the synchronous gauge. Once the synchronous gauge is properly fixed, one finds that the induced GW spectrum in the synchronous gauge and in the Newtonian gauge yields the same prediction in a radiation dominated universe \cite{Inomata:2019yww}. Later on, Ref. \cite{Domenech:2020xin} proposed that under certain conditions the energy density of SIGWs can be gauge independent up to $\mathcal{O}(\mathcal{H}^2/k^2)$ corrections. 

Unfortunately, the above approaches do not work well in our scenario. In the SIGWs case people are concerned about observables that can be measured by detectors, so preferred gauges can be chosen along this line. In our scenario, we are interested in the self-consistency of matter-bounce scenario. At late times, the SIVPs will be diluted by cosmic expansions. Thus, the overproduction of SIVPs is not possible to be connected to any observables measurable at present. The SIGWs are measures at subhorizon regions where cosmology is less relevant, while the overproduction of SIVPs is relevant to the superhorizon modes. The $\mathcal{O}(\mathcal{H}^2/k^2)$ corrections are tolerable in SIGWs but important in SIVPs. As we have already seen in the above section, the different resultants spectra index and $c_s$ dependence in UFG and CCG originate from the $\mathcal{H}^2/k^2$ factor in the Einstein equations. 

Thus, we will interpret our result as follows:
\begin{itemize}
	\item From the theoretical perspective, we are interested in the backreaction of secondary fluctuations. The ultimate goal is to prove that energy density of secondary fluctuations can be as large as the background one in matter-bounce cosmology. 
	\item Vector fluctuations grow as the universe contracts. Intuitively, given a certain spacetime foliation, the corresponding vector modes are more likely to be problematic, which is the basic motivation of our work. The energy density of secondary vectors $\rho_V^{(2)}$ is apparently smaller than the energy density of secondary fluctuations $\rho^{(2)}$. We find that $\rho_V^{(2)}$ is much larger than $\rho_{\rm bg}$ in both UFG and CCG, once the curvature fluctuations match the CMB observations. Thus, the energy density of the secondary fluctuations is larger than that of the background, indicating a backreaction problem.
	\item From the observational perspective, an observable measured by detectors shall be gauge invariant. The SIVPs, as discussed above, is not such an observable and we cannot pick a preferred gauge that is favored by observations. 
	
	\item However, we note that the action \eqref{eq:kessence} is an effective description of the contraction phase. By adopting \eqref{eq:kessence}, one assumes that other fields are subdominant compared to the k-essence field. Therefore, the overproduction of SIVPs indicates that the above effective description is broken down. One has to consider the presence of fields that contain vector degree of freedoms in the contraction phase. 
	
	\item As long as the additional field(s) are added to the action, we can talk about their energy density as observables. Vector fluctuations can no longer be taken as secondary due to the presence of vector degrees of freedom. Instead, we shall treat the vector field(s) at the matter sector and the vector fluctuations at the geometry sector as being at a linear level. Now, the energy density of the vectors can be properly defined thanks to the SVT decomposition. Generically, the energy density of vectors at linear level is much larger than the nonlinear level, and the well-defined linear energy density is sufficient to be confronted with observations.
	
	\item When the secondary energy density of the vectors exceeds the linear one, we will potentially face a similar gauge issue of vector modes. In this case, we have to specify the detailed vector field(s) and discuss how it is to be measured by astrophysical observations. After that, we can pick up a preferred gauge using a similar method to the SIGWs studies. Unfortunately, this approach is highly model-dependent and far beyond the scope of the current work.
\end{itemize}

\section{Energy Density of SIVPs}

The energy density of the SIVPs is given by~\cite{Ota:2021fdv}
\begin{equation} \label{def:V}
    \rho_V(\tau,\vec{x}) = \frac{1}{4a^2} \partial_i G_j(\tau,\vec{x}) \partial^i G^j(\tau,\vec{x}) ~.
\end{equation}
From Eq.~\eqref{def:V}, the energy density of SIVPs is related to $\mathcal{P}_G$ as
\begin{equation}
\rho_V (\tau) = \frac{1}{2a^2} \int \ddd k k \mathcal{P}_G (\tau,\vec{k}) ~.
\end{equation}
The backreaction is represented by the ratio of the energy density of the SIVPs against the background one at $\tau = \tau_0$:
\begin{equation}
    \label{eq:deltaVdef}
    \delta_V \equiv \frac{\rho_V(\tau_0)}{\rho_{\rm bg} (\tau_0)} = \frac{1}{24} \int (k\tau_0) \mathcal{P}_G(k,\tau_0) \ddd(k\tau_0) ~.
\end{equation}

The $\mathcal{P}_{\zeta}$ on superhorizon scales is associated to CMB observations:
\begin{equation}
\label{eq:PzetaAs}
    \mathcal{P}_{\zeta}(\vec{k},\tau_0) \simeq \frac{1}{12\pi^2 c_s \tau_0^2}  = A_s ~,
\end{equation}
where $A_s = 2.1 \times 10^{-9}$ from the Planck collaboration \cite{Planck:2018vyg}. We denote the scales of fluctuations that ``cross the horizon'' at the beginning/end of the contraction phase to be $k_{\rm min}$ and $k_{\rm \max}$: 
\begin{align}
    k_{\rm min} &= c_s^{-1}|\mathcal{H}(\tau_{\rm ini})| = -2(c_s \tau_{\rm ini})^{-1} ~,
    \\
    k_{\rm max} &= c_s^{-1}|\mathcal{H}(\tau_0)| = -2(c_s \tau_0)^{-1} ~,
\end{align}
where $\tau_{\rm ini}$ labels the initial time of matter-contraction phase. The modes with $k_{\rm min} < k < k_{\rm \max}$ become superhorizon during the matter-contraction phase, and we adopt a minimal curvature power spectrum for superhorizon perturbations
\begin{equation}
\label{eq:Pzetaansatz}
    \mathcal{P}_{\zeta} = \begin{cases}
        A_s ~,~ & k_{\rm min} < k < k_{\rm \max} ~, \\
        0 ~,~ & \text{otherwise}~.
    \end{cases}
\end{equation}
It is possible that modes with $k < k_{\rm min}$ or $k > k_{\rm \max}$ become superhorizon before or after the matter-contraction phase, hence they also give a positive contribution to SIVPs. Nonetheless, we have no knowledge on the dynamics of these modes in matter-contraction phase without information about precontraction and postcontraction phases. We thus adopt the ansatz in Eq.~\eqref{eq:Pzetaansatz}, which captures the dominant contribution to SIVPs from modes entering the horizon during the contraction phase, providing a lower bound sufficient to analyze the SIVPs instability.

The scale of scale-invariant curvature fluctuation indicated by CMB observations ranges from $k_C/a_{\rm today} \simeq 10^{-4}~\textnormal{Mpc}^{-1}$ to $k_L/a_{\rm today} \simeq 1~\textnormal{Mpc}^{-1}$.  To match CMB data, modes with $k = k_C$ must be well within the horizon ($k_C \gg k_{\rm min}$), while $k_L$ must be superhorizon at $\tau = \tau_0$ ($k_L \leq k_{\rm max}$). We introduce a dimensionless scaling factor
\begin{equation} \label{def:g}
   g \equiv \frac{\tau_{\rm ini} }{ \tau_0} = \frac{ k_{\rm max}}{k_{\rm min}} \gg \frac{k_L}{k_C} = 10^4 ~, 
\end{equation}
to label the duration of contraction phase.

\begin{figure}[ht]
    \centering
    \includegraphics[width=0.45\textwidth]{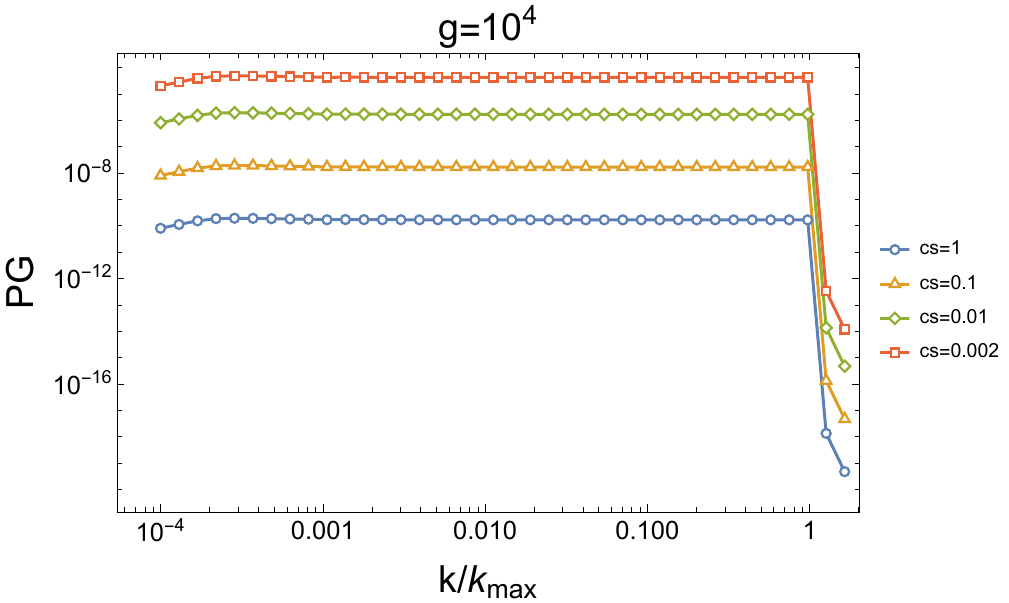}  
    \includegraphics[width=0.45\textwidth]{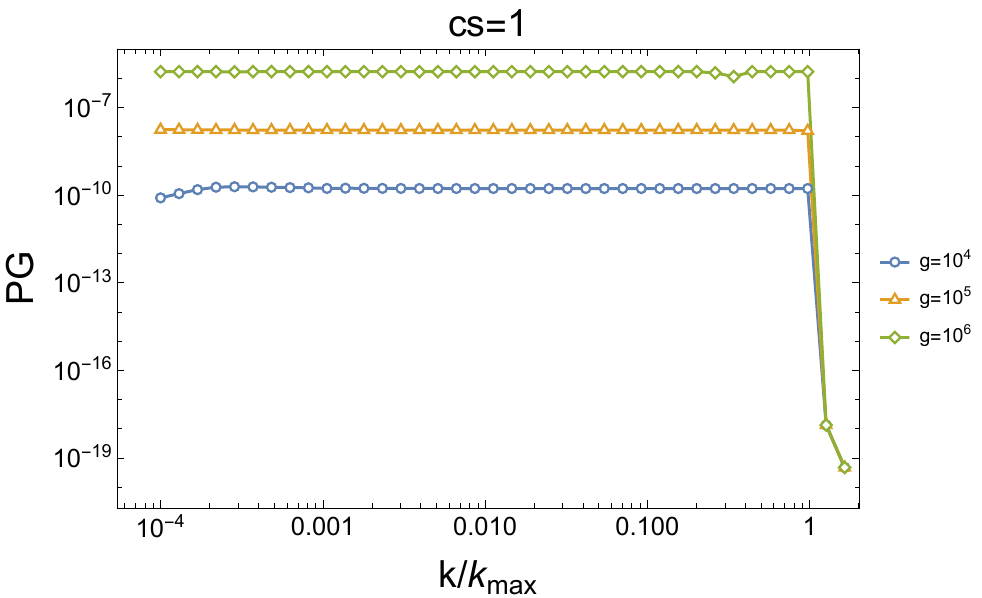}
    \caption{Top: SIVP power spectra~\eqref{eq:PG} at the end of matter-contraction $\tau_0$ as functions of $c_s$ with $g = 10^4$. Bottom: The SIVP power spectra~\eqref{eq:PG} as functions of $g$ with $c_s = 1$.}
    \label{fig:PGmaintext}
\end{figure}

The resulting vector power spectrum is determined by the dimensionless parameters $c_s$ and $g$, as $\tau_0$ is fixed by $c_s$ through Eq.~\eqref{eq:PzetaAs}. We numerically evaluate $\mathcal{P}_G$ in the uniform field gauge, with the results presented in Fig.~\ref{fig:PGmaintext}.
We demonstrate both numerically and analytically that $\mathcal{P}_G$ scales as $A_s^2 c_s^{-2} g^2$, with detailed derivations provided in Appendix~\ref{app:num}. Consequently,
\begin{equation}
    \delta_V(\tau_0) = \mathcal{C} A_s^2 \frac{g^2}{c_s^4} ~,
\end{equation}
where the structural constant is $\mathcal{C} \simeq 3.1 \times 10^{-2}$ according to the numerical results. 

For comparison, the energy density of linear curvature perturbations scales as $\ln g$ assuming a scale-invariant $\mathcal{P}_{\zeta}$ and the fact of the dominance of kinetic energy due to the superhorizon growth of curvature fluctuation. As a result, the backreaction problem associated with SIVPs can be more severe than that of curvature fluctuations due to its $g^2 c_s^{-4}$ scaling. 

The value of $\delta_V$ is presented in Table \ref{tab:PGUFG}. Considering the consistency relation in the context of k-essence theory during matter-contraction, we have $r = 24 c_s$~\cite{Li:2016xjb}, where $r$ is the tensor-to-scalar ratio constrained by $r_{0.002} < 0.044$~\cite{Planck:2018jri}. This constraint implies $c_s < 0.002$, leading to an excessively large $\delta_V$ along with the fact $g \gg 10^4$. The marginally accepted sound speed $c_s = 0.002$ with $g = 10^4$ already leads to $\delta_V = 0.85$, namely the energy density of SIVPs becomes comparable to the background one, indicating either a significant backreaction on the background evolution or a breakdown of perturbation theory.
 
The situation becomes worse once we consider the energy density of SIVPs versus that of curvature fluctuations. Secondary fluctuations, by definition, shall have a much lower energy density than that of linear fluctuations. We compute the ratio of energy density of SIVPs versus $\zeta$ as \cite{supp}
\begin{equation}
    \frac{\rho_V(\tau_0)}{\rho_{\zeta}(\tau_0)} \simeq \delta_V(\tau_0) \frac{\rho_{\rm bg}(\tau_0)}{\rho_{\zeta}(\tau_0)} = \frac{8}{3}\mathcal{C} \frac{A_s}{c_s^4} \frac{g^2}{\ln g} ~,
\end{equation}
which already acquires a considerably large value $1.9 \times 10^{-3}$ when taking $g = 10^4$ and $c_s = 1$. The restriction $c_s < 0.02$ will lead to $\rho_V(\tau_0)/\rho_{\zeta}(\tau_0) > 1$ and the breakdown of perturbation theory. 

We also compute the SIVPs in the constant curvature gauge and find out the corresponding scaling $\delta_V \propto A_s^2g^2c_s^{-8}$, as already worked out in Sec. IV D. Thus, in this gauge, the SIVPs possess a more severe challenge than in the uniform field gauge due to the restriction $c_s < 0.002$. As SIVPs overproduction is found in both gauges, we conclude that the vector instability is a generic fact rather than a purely gauge defect.

Combining these arguments, we conclude that cosmological models where nearly scale-invariant curvature fluctuations on CMB and LSS scales originate from a matter-contraction phase governed by a minimally coupled k-essence field are constrained by the overproduction of SIVPs, rendering such models invalid.

\section{Conclusion}

Vector fluctuations play a crucial role in bouncing cosmologies, particularly during the contraction phase. This study, for the first time, combines the concept of secondary vector fluctuations induced by scalar fluctuations with observational constraints on curvature perturbations, revealing an overproduction of these modes in a matter-contraction phase governed by a k-essence scalar field. This finding highlights the importance of vector modes in bouncing cosmologies and motivates further investigation into their impact on various bouncing scenarios, including matter-bounces with more complex actions~\cite{Akama:2019qeh}, Ekpyrotic scenarios~\cite{Khoury:2001wf}, and scenarios where the bouncing phase significantly influences the evolution of curvature fluctuations~\cite{Cai:2012va}.

Our findings motivate further exploration of perturbation theory within the context of bouncing cosmology. For instance, our results could be revisited by replacing the cutoff of the curvature power spectrum with regularized primordial fluctuations. This approach could, in principle, yield more accurate results. However, this area currently lacks sufficient research. The smallness of sound speed $c_s \ll 1$ could also lead to oversized non-Gaussianities \cite{Li:2016xjb}, and it is worthy studying if there are relations between the SIVPs and the non-Gaussianities, especially the scalar trispectrum. Additionally, the growth of anisotropic shear during the contraction phase, whose energy density scales as $\rho \propto a^{-6}$, warrants investigation of scalar-induced shear, which could provide additional theoretical constraints on the bouncing models. The gauge dependence of SIVPs, much less explored compared to the SIGW case, should be investigated in more detail. Finally, the secondary vector fluctuations in bouncing cosmology in the presence of vector field(s) deserve future investigation due to their potential connection with topics such as primordial magnetogenesis ~\cite{Kandus:2010nw}. \\

\appendix

\begin{widetext}

\section{Vector Power Spectrum from Numerical Evaluation}
\label{app:num}
We are interested in the energy density of the induced vector perturbation,
\begin{equation}
	\rho_V(\vec{x},\tau) = \frac{1}{4a^2} \partial_i G_j(\vec{x},\tau) \partial^i G^j (\vec{x},\tau) ~,
\end{equation}
which is related to the vector power spectrum as 
\begin{equation}
	\rho_V (\tau) = 2 \times \frac{1}{4a^2} \int k \mathcal{P}_G (\tau,\vec{k}) \ddd k = \frac{1}{2a^2} \int k \mathcal{P}_G (\tau,\vec{k}) \ddd k ~.
\end{equation}
The factor of 2 arises from the two polarizations of vector perturbations. The vector energy density, $\rho_V$, is directly determined by the vector power spectrum, $\mathcal{P}_G$. 

Once we work out the time integral, we may compute the vector power spectrum numerically using \eqref{eq:PGbare} and the curvature power spectrum
\begin{equation}
	\label{eq:Pzetanum}
	\mathcal{P}_{\zeta} = \begin{cases}
		A_s ~,~ & k_{\rm min} < k < k_{\rm \max} ~, \\
		0 ~,~ & \text{otherwise} ~.
	\end{cases}
\end{equation}
For convenience, we will also introduce a dimensionless scaling factor
\begin{equation} 
	g \equiv \frac{\tau_{\rm ini} }{ \tau_0} = \frac{ k_{\rm max}}{k_{\rm min}} ~, 
\end{equation}
which labels the duration of the contraction phase. The scale of scale-invariant curvature fluctuation indicated by CMB observations ranges from $k_C/a_{\rm today} \simeq 10^{-4}~\textnormal{Mpc}^{-1}$ to $k_L/a_{\rm today} \simeq 1~\textnormal{Mpc}^{-1}$.  To match CMB data, the modes with $k = k_C$ must be well within the horizon ($k_C \gg k_{\rm min}$), while $k_L$ must be superhorizon at $\tau = \tau_0$ ($k_L \leq k_{\rm max}$). Therefore
\begin{equation}
	g = \frac{ k_{\rm max}}{k_{\rm min}} \gg \frac{k_L}{k_C} = 10^4 ~.
\end{equation}

\begin{figure}[ht]
	\centering
	\includegraphics[width=0.5\textwidth]{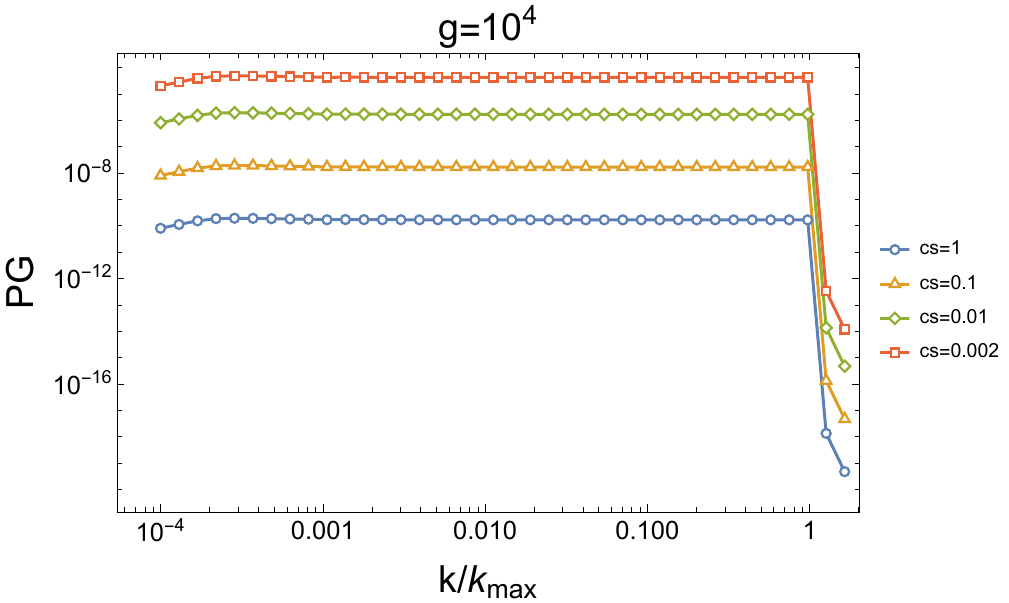} \\
	\includegraphics[width=0.5\textwidth]{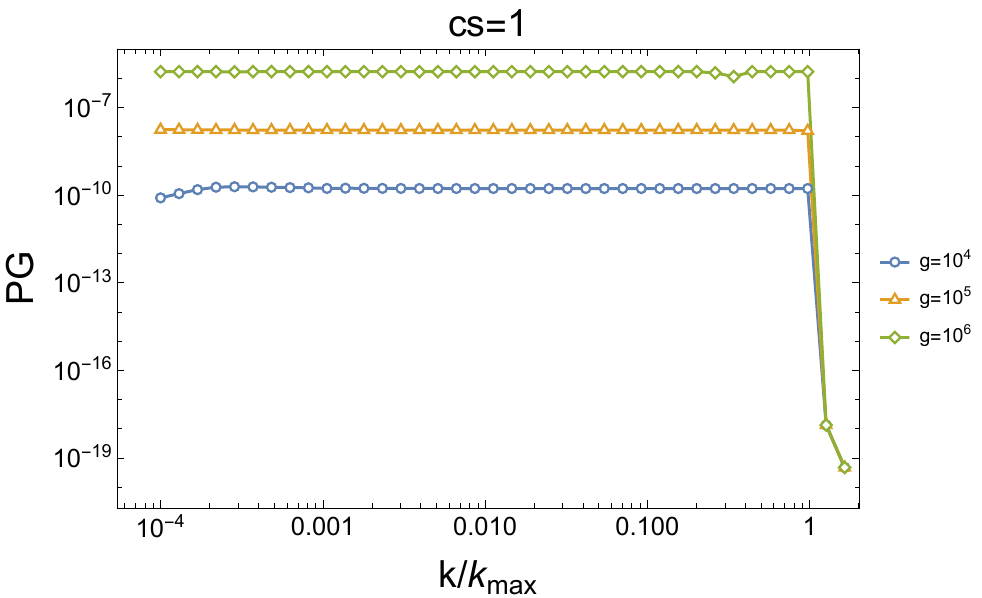}
	\caption{The vector power spectrum $\mathcal{P}_G$ calculated using Eqs. \eqref{eq:PGbare} and \eqref{eq:Pzetanum}. We show $\mathcal{P}_G$ for different $c_s$ with $g = 10^4$ fixed in the upper panel, and $\mathcal{P}_G$ for different $g$ with $c_s = 1$ fixed in the lower panel. }
	\label{fig:PG}
\end{figure}

We organize the resultant $\mathcal{P}_G$ as a function of $c_s$ and $g$ in Fig. \ref{fig:PG}. We see that $\mathcal{P}_G$ is nearly scale invariant on scales $k < k_{\rm max}$. We schematically parametrize it as a power-law function
\begin{equation}
	\label{eq:PGpara}
	\mathcal{P}_G(\tau_0,k) = \mathcal{A}_G \left( \frac{k}{k_{\rm max}} \right)^{n_G} ~,~ k < k_{\rm max} ~.
\end{equation}
We thus obtain the ratio of $\rho_V$ versus the background energy density value $\rho_{\rm bg}$,
\begin{equation}
	\delta_V \equiv \frac{\rho_V(\tau_0)}{\rho_{\rm bg} (\tau_0)} = \frac{1}{24} \int (k\tau_0) \mathcal{P}_G(k,\tau_0) \ddd(k\tau_0) ~,
\end{equation}
as
\begin{align}
	& \quad \delta_V(\tau_0) \nonumber = \frac{1}{24} \int (k\tau_0) \mathcal{P}_G(k,\tau_0) \ddd(k\tau_0) \\
	& < \frac{1}{6c_s^2} \int_0^1 \mathcal{A}_G \left( \frac{k}{k_{\rm max}} \right)^{n_G+1} \ddd \left( \frac{k}{k_{\rm max}} \right) = \frac{\mathcal{A}_G}{6c_s^2 (n_G + 2)} \simeq \frac{\mathcal{A}_G}{12c_s^2} ~.
\end{align}
We note that the value of $\delta_V(\tau_0)$ is insensitive to the lower limit of the integral. So, \eqref{eq:PGpara} is sufficient to evaluate $\delta_V(\tau_0)$ although it could be invalid in the limit $k \to 0$. We organize the vector power spectrum for different model parameters in Table \ref{tab:PGUFG}. 
\begin{table}[ht]
	\centering
	\begin{tabular}{|c|c|c|c|c|}
		\hline
		\multicolumn{2}{|c|}{Model parameters} & \multicolumn{3}{|c|}{Outcomes} \\
		\hline 
		$c_s$ & $g$ & $\mathcal{A}_G$ & $n_G$ & $\delta_V (\tau_0)$ \\ 
		\hline
		$2 \times 10^{-3}$ & $10^4$ & $4.1 \times 10^{-5}$ & $0.01$ & $8.5 \times 10^{-1}$ \\  
		\hline
		$10^{-2}$ & $10^4$ & $1.6 \times 10^{-6}$ & $0.01$ & $1.3 \times 10^{-3}$ \\  
		\hline
		$10^{-1}$ & $10^4$ & $1.6\times 10^{-8}$ & $0.01$ & $1.3 \times 10^{-7}$ \\ 
		\hline
		$1$ & $10^4$ & $1.6 \times 10^{-10}$ & $0.02$ & $1.3 \times 10^{-11}$ \\
		\hline
		$1$ & $10^5$ & $1.6 \times 10^{-8}$ & $0.02$ & $1.3 \times 10^{-9}$ \\
		\hline
		$1$ & $10^6$ & $1.6 \times 10^{-6}$ & $0.02$ & $1.3 \times 10^{-7}$ \\
		\hline
	\end{tabular}
	\caption{Vector power spectra \eqref{eq:PGpara} and the values of $\delta_V$ for various values of $c_s$ and $g$.}
	\label{tab:PGUFG}
\end{table}	
It is easy to see that $\mathcal{A}_G$ scales as $c_s^{-2}g^2$. Accordingly, $\delta_V(\tau_0)$ has the schematic form
\begin{equation}
	\delta_V(\tau_0) = \mathcal{C} A_s^2 \frac{g^2}{c_s^4} ~,
\end{equation}
and the numerical result tells us that the structural constant is $\mathcal{C} \simeq 3.1 \times 10^{-2}$.

In the end, let us compare the energy density of SIVPs versus that of the curvature fluctuation. The sound speed is restricted by $c_s < 0.002$ and $\delta_V \simeq 0.85$, which already sets a serious problem. The issue becomes more severe if we consider the restriction from curvature energy density. On a superhorizon scale,
\begin{equation}
	(\zeta^{\prime})^2 \sim \frac{9}{4} \mathcal{H}^2 \zeta_k^2 \gg c_s^2(\partial \zeta)^2 \sim c_s^2k^2 \zeta_k^2 ~.
\end{equation}
Thus, we may estimate the energy density of curvature fluctuation using its kinetic energy
\begin{equation}
	\rho_{\zeta}(\tau_0) \sim \frac{1}{2} \int_{k_{\rm min}}^{k_{\rm max}} (\zeta_k^{\prime})^2 d\ln k \sim \frac{9}{8} \mathcal{H}(\tau_0)^2 \mathcal{P}_{\zeta} \ln g ~,
\end{equation}
which gives
\begin{equation}
	\frac{\rho_V(\tau_0)}{\rho_{\zeta}(\tau_0)} \simeq \delta_V(\tau_0) \frac{\rho_{\rm bg}(\tau_0)}{\rho_{\zeta}(\tau_0)} = \frac{8}{3}\mathcal{C} \frac{A_s}{c_s^4} \frac{g^2}{\ln g} ~,
\end{equation}
and its minimum value, when taking $g = 10^4$ and $c_s = 1$, is already $1.9 \times 10^{-3}$. Therefore, the restrictions $c_s < 2.0 \times 10^{-3}$ and $g \gg 10^4$ will lead to $\rho_V(\tau_0)/\rho_{\zeta}(\tau_0) \gg 1$. The energy density of secondary perturbations greatly exceeds that of the linear one, which could imply the breakdown of perturbation theory.

We acknowledge the usage of MathGR~\cite{Wang:2013mea} in the project.

\end{widetext}

\section*{Acknowledgments}

We thank Yi Wang for discussions which inspired this work and Xian Gao, Chunshan Lin and George Zahariade for their critical comments. We also thank Shingo Akama, Yi-Fu Cai, Yong Cai, Alexander Ganz, and Sabino Matarrese for the discussions on several technical problems. We thank the anonymous referees whose comments on the gauge problems greatly helped to improve the quality of the work. We are grateful to Atsuhisa Ota for his contributions during the initial stage of this work. M.Z. was supported by grant No. UMO 2021/42/E/ST9/00260 from the National Science Centre, Poland.
C.C. is grateful for the support from the Jockey Club Institute for Advanced Study at The Hong Kong University of Science and Technology and is supported by NSFC Grants No. 12433002.

\bibliographystyle{utphys}
\bibliography{siv}
\end{document}